\documentclass[12pt]{article}
\pdfoutput=1
\usepackage{pstricks}
\usepackage{color}
\usepackage{amssymb,amsmath,bm,bbm}
\usepackage{epsf}
\usepackage{epsfig}
\usepackage{afterpage}
\usepackage{longtable}
\usepackage{cite}
\usepackage{latexsym,mathrsfs,dsfont}
\usepackage{graphics}
\usepackage{url}
\usepackage{paralist}
\usepackage{bbold}

\newcommand{\svs}{\vbox{\vskip 5mm}}

\newcommand{\ord}{\mathcal{O}}

\newcommand{\IM}{\rm{Im}}
\newcommand{\RE}{\rm{Re}}

\newcommand{\tev}{\, {\rm TeV}}
\newcommand{\gev}{\, {\rm GeV}}
\newcommand{\mev}{\, {\rm MeV}}

\newcommand{\vcb}{|V_{cb}|}

\newcommand{\vub}{|V_{ub}|}

\newcommand{\bsi}{B_6^{(1/2)}}
\newcommand{\bei}{B_8^{(3/2)}}

\def\epe{\varepsilon'/\varepsilon}
\newcommand{\beq}{\begin{equation}}
\newcommand{\eeq}{\end{equation}}
\newcommand{\be}{\begin{equation}}
\newcommand{\ee}{\end{equation}}
\newcommand{\bi}{\begin{itemize}}
\newcommand{\ei}{\end{itemize}}
\newcommand{\ba}{\begin{array}}
\newcommand{\ea}{\end{array}}
\newcommand{\beqa}{\begin{eqnarray}}
\newcommand{\eeqa}{\end{eqnarray}}
\newcommand{\bea}{\begin{eqnarray}}
\newcommand{\eea}{\end{eqnarray}}
\newcommand{\beqn}{\begin{eqnarray}}
\newcommand{\eeqn}{\end{eqnarray}}

\newcommand{\D}{\Delta}

\definecolor{red}{cmyk}{0,1,1,0.4}

\def\kpn{K^+\rightarrow\pi^+\nu\bar\nu}
\def\klpn{K_{L}\rightarrow\pi^0\nu\bar\nu}

\usepackage{fancyhdr}
\advance \headheight by 3.0truept       

\begin{document}

\begin{flushright}
    {FLAVOUR(267104)-ERC-70}\\
    {BARI-TH/14-689}
\end{flushright}

\medskip

\begin{center}
{\LARGE\bf
\boldmath{$Z$-$Z^\prime$ Mixing and $Z$-Mediated FCNCs\\ in 
 $SU(3)_C\times SU(3)_L\times U(1)_X$ Models}}
\\[0.8 cm]
{\bf Andrzej~J.~Buras$^{a,b}$, Fulvia~De~Fazio$^{c}$ and
Jennifer Girrbach-Noe$^{a,b}$
 \\[0.5 cm]}
{\small
$^a$TUM Institute for Advanced Study, Lichtenbergstr. 2a, D-85747 Garching, Germany\\
$^b$Physik Department, Technische Universit\"at M\"unchen,
James-Franck-Stra{\ss}e, \\D-85747 Garching, Germany\\
$^c$Istituto Nazionale di Fisica Nucleare, Sezione di Bari, Via Orabona 4,
I-70126 Bari, Italy}
\end{center}

\vskip0.41cm


\abstract{%
\noindent
Most of the existing analyses of flavour changing neutral current processes (FCNC) in 
the 331 models,  based on the gauge group $SU(3)_C\times SU(3)_L\times U(1)_X$, 
are fully dominated by tree-level exchanges of a new heavy neutral gauge 
boson $Z^\prime$. However, due to the $Z-Z^\prime$ mixing also 
corresponding contributions from $Z$ boson are present. As the $Z-Z^\prime$ 
mixing is estimated generally in $Z^\prime$ models to be at most $\ord(10^{-3})$,  the latter contributions are usually neglected. The paucity of relevant 
parameters in 331 models allows to check whether this neglect is really 
justified in these concrete models. We calculate the impact of these contributions on $\Delta F=2$ processes and rare $K$, $B_s$ and $B_d$ decays for 
 different values of a parameter $\beta$, which distinguishes between various 331 models and for  different fermion representations under the $SU(3)_L$ group.
 We find  a general expression for the $Z-Z^\prime$ mixing in terms 
 $\beta$, $M_Z$, $M_{Z^\prime}$ and $\tan\bar\beta$, familiar from 2 Higgs Doublet models, that differs from the one quoted in the 
literature.
 We study in particular the models with $\beta=\pm n/\sqrt{3}$ with $n=1,2$ 
which have recently been investigated by us in the context of new data 
 on $B_{s,d}\to \mu^+\mu^-$ and $B_d\to K^*(K)\mu^+\mu^-$. We find that 
these new contributions can indeed be neglected in the case of 
$\Delta F=2$ transitions and decays, like  $B_d\to K^*\mu^+\mu^-$, where they are suppressed by the small vectorial $Z$ coupling to charged 
leptons. However, 
the contributions of tree-level $Z$ exchanges to decays sensitive to 
axial-vector couplings, like $B_{s,d}\to \mu^+\mu^-$ and $B_d\to K\mu^+\mu^-$, 
and those with neutrinos in the final state, like  $b\to s\nu\bar\nu$ transitions, $\kpn$ and $\klpn$ cannot be  generally neglected with size of 
$Z$ contributions depending on $\beta$, $\tan\bar\beta$ and $M_{Z^\prime}$. We analyze how our recent results 
on FCNCs in 331 models, in particular correlations between various observables, are modified by these new contributions.  As a byproduct we 
analyze for the first time the ratio $\epe$ in 
these models including both $Z^\prime$ and $Z$ contributions.   Our 
analysis of electroweak precision  observables within 331 models demonstrates transparently 
that the interplay of NP effects in electroweak precision observables and those in flavour observables could allow in the future to identify the favourite 
331 model.}

\thispagestyle{empty}
\newpage
\setcounter{page}{1}

\tableofcontents

\section{Introduction}
An interesting class of dynamical models  are
the 331 models based on the gauge group $SU(3)_C\times SU(3)_L\times U(1)_X$
  \cite{Pisano:1991ee,Frampton:1992wt}. Detailed analyses of FCNC processes
in these models have been presented by us in \cite{Buras:2012dp,Buras:2013dea}. Selection of earlier analyses of various aspects of these models
related to flavour physics
can be found in
\cite{Ng:1992st,Diaz:2003dk,Liu:1993gy,Diaz:2004fs,Ochoa:2005ih,Liu:1994rx,Rodriguez:2004mw,Promberger:2007py,Agrawal:1995vp,CarcamoHernandez:2005ka,Promberger:2008xg}. For other variants of 331 models see \cite{Singer:1980sw,Valle:1983dk,Boucenna:2014ela}.
{ We briefly recall here only a few aspects of these models relevant for the analysis in this paper.
As will be discussed in section  \ref{Frepresentation}, fermion representations under $SU(3)_L$ transformations can be chosen in several ways. However, requirement of anomaly cancelation and asymptotic freedom of QCD imposes that, for example, if two quark generations transform as triplets, the remaining one should be an antitriplet. 
An interesting relation connects the electric charge $Q$ to the  generators $T_3,\,T_8$ of $SU(3)$ and the generator $X$ of $U(1)_X$: $Q=T_3 + \beta T_8+X$, introducing the parameter $\beta$ that plays a key role in this class of models. 

Having an enlarged gauge group with respect to the SM, a number of new gauge bosons is present, whose charge depend on the value of $\beta$.   However, independently of it,  a new neutral gauge boson $Z^\prime$ is always present  and can mediate FCNC at tree level in the quark sector. The Higgs sector is also enlarged. In particular, three Higgs triplets are present. Among these, two give masses to up and down type quarks, and the relative size of their vacuum expectation values will be important for our subsequent discussion. We shall introduce them later in Section \ref{sec:2}. Finally, also new heavy fermions are predicted to exist, but they do not play any role in our study and we shall not consider them any longer.}

The nice feature of these models is a small number of free parameters which
is lower than present in general $Z^\prime$  scenarios with left-handed 
flavour violating couplings to quarks considered
in \cite{Buras:2012jb,Buras:2013qja}. This allows to find certain
correlations between different  meson systems which is not possible in the general case. Indeed
the strength of the relevant $Z^\prime$ couplings  to down-quarks is governed 
 in these models by two mixing parameters, two CP-violating phases and the
parameter $\beta$ which 
defines a given 331 model up to the choice of fermion representations 
 \cite{Diaz:2004fs,CarcamoHernandez:2005ka} and
determines the charges of new heavy fermions and gauge bosons { as we have already mentioned above}.

Thus for a given $M_{Z^\prime}$ and $\beta$ there are only four free parameters to our disposal. In particular for a given $\beta$, the diagonal couplings of $Z^\prime$ to quarks and leptons are fixed. Knowing these
  couplings simplifies the analysis significantly, increasing simultaneously the predictive power of the theory.

In \cite{Buras:2012dp} the relevant couplings have been presented for arbitrary  $\beta$ and in \cite{Buras:2013dea} a particular set of models with 
\be\label{nbeta}
\beta=\pm \frac{n}{\sqrt{3}}, \qquad n=1,2,3
\ee 
has been analyzed. We have demonstrated that
\begin{itemize}
\item
The models with $\beta=-2/\sqrt{3}$ and $\beta=-1/\sqrt{3}$ help in understanding  the anomalies in $B_d\to K^*\mu^+\mu^-$  \cite{Aaij:2013iag,Aaij:2013qta} because in these models the coupling $\Delta_V^{\mu\bar\mu}(Z^\prime)$  is large. 
\item
The models with $\beta=2/\sqrt{3}$ and $\beta=1/\sqrt{3}$ having significant 
$\Delta_A^{\mu\bar\mu}(Z^\prime)$ coupling
provide interesting 
NP effects in  $B_{s}\to\mu^+\mu^-$ that allow to bring the theory 
closer to the data \cite{Aaij:2013aka,Chatrchyan:2013bka,CMS-PAS-BPH-13-007}.
\item
The model with $\beta=-\sqrt{3}$, advocated in particular in  \cite{Gauld:2013qja} in 
the context of $B_d\to K^*\mu^+\mu^-$ anomalies, has several problems 
originating in the presence of  a Landau singularity for $\sin^2\theta_W=0.25$. { The same problem is found in the case $\beta=\sqrt{3}$}.
Therefore we will not consider  them here.
\end{itemize} 

 We would like to emphasize here that these results have been obtained 
by assigning the fermions to specific fermion representations under 
$SU(3)_L$ and that even for a given $\beta$ the results listed above 
can change if the choice of representations is different. While it is 
known that the phenomenology of 331 models depends on the choice 
of fermion representations \cite{Diaz:2004fs,CarcamoHernandez:2005ka} we recall some 
aspects of it below as this freedom has interesting consequences in 
the context of our analysis. Moreover, it clarifies certain differences 
between the analyses in \cite{Diaz:2004fs,CarcamoHernandez:2005ka} and 
\cite{Buras:2012dp,Buras:2013dea}\footnote{We thank R. Martinez and F. Ochoa 
for discussions on this point.}.

However, our previous analyses and to our knowledge all analyses of FCNC processes 
in 331 models  neglected contributions from tree-level $Z$ boson exchanges. 
Such contributions can be generated in 331 models by the $Z-Z^\prime$ mixing 
but were expected to be very small
as according to general analyses
 \cite{Langacker:2008yv,Erler:2009jh} this mixing should be at most of $\ord(10^{-3})$. In the absence of  this mixing the $Z$ couplings remain for a given 
electric quark charge flavour universal and
in contrast to $Z^\prime$ gauge boson, there are no FCNCs mediated by $Z$  boson at tree-level.

The goal of the present paper is to investigate, whether the neglect 
of $Z$ boson FCNCs generated by $Z-Z^\prime$ mixing in the 331 
models in question is really justified. It will turn out 
that this is not always the case and we will investigate what is the impact 
of these new contributions on our results in  \cite{Buras:2013dea}. Fortunately 
it will turn out that the determination of the allowed ranges for the 
parameters of the 331 models through $\Delta F=2$ processes is unaffected 
by these new contributions. The same applies to our analysis of the anomalies in $B_d\to K^*\mu^+\mu^-$.  On the other hand in other decays considered 
by us $Z$ contributions can be as large as $Z^\prime$ contributions so that 
for certain parameters and models the two contributions can cancel each other.

At this point we would like to emphasize that even if 331 models would 
not survive future flavour precision tests, they offer a very powerful 
laboratory to study not only correlations between various 
flavour observables but also between flavour observables and electroweak 
precision observables. One of the goals of our paper is to exhibit these 
correlations transparently.

Our paper is organized as follows. In Section~\ref{sec:2} we summarize 
 some aspects of 331 models and 
present the general expression for the $Z-Z^\prime$ mixing that differs 
from the one quoted in the literature \cite{Diaz:2004fs,CarcamoHernandez:2005ka}.
We also  analyze 
for which processes and for which values of $\beta$ the resulting FCNCs
mediated by $Z$ boson are relevant. We will frequently refer to our 
previous papers  \cite{Buras:2012dp,Buras:2013dea}, where all the details 
on the models considered can be found. In particular in the Appendix A in 
\cite{Buras:2013dea} a compendium of all $Z^\prime$ couplings and $Z$ couplings 
including their numerical values can be found. We will not repeat this 
compendium here but we will use it extensively in order to find out 
already in this section where the neglect of flavour violating $Z$ exchanges is justified and where they have to be taken into account. 
In Section~\ref{sec:3} 
we show how  our results in \cite{Buras:2012dp,Buras:2013dea} 
are modified through the inclusion of $Z$ 
contributions. In Section~\ref{sec:4} we present for the first time the 
analysis of $\epe$ in 331 models and its correlation with rare $K$ decays. 
In Section~\ref{EWP} 
we reconsider the effects of $Z-Z^\prime$ mixing in electroweak precision 
observables and discuss correlations between flavour and electroweak 
precision observables in 331 models in question.
 We conclude in  Section~\ref{sec:5}.

\boldmath
\section{$Z-Z^\prime$ Mixing in 331 Models}\label{sec:2}
\unboldmath
\subsection{Basic Formulae for  $Z-Z^\prime$ Mixing}
Among the new heavy particles in 331 models the most important role in flavour 
physics is played 
by a new $Z^\prime$ boson originating in the additional $U(1)_X$ factor in 
the extended gauge group. The electroweak symmetry breaking is discussed 
in several papers quoted above and we will not repeat it here. It suffices to state that 
after the mass eigenstates for the SM fields, the photon and the $Z$ boson 
have been constructed through appropriate rotation, there remains still 
small mixing between $Z$ and $Z^\prime$ so that the heavy mass eigenstates 
are really 
\be
Z_\mu^1=\cos\xi Z_\mu + \sin\xi Z^\prime_\mu, \qquad 
Z_\mu^2=-\sin\xi Z_\mu + \cos\xi Z^\prime_\mu \, .
\ee
As $\sin\xi$ is estimated to be at most $\ord(10^{-3})$ this mixing is usually 
  neglected in FCNC processes so that the two mass eigenstates are simply $Z_\mu^1=Z_\mu$ and 
$Z_\mu^2=Z^\prime_\mu$. Consequently only $Z^\prime_\mu$ has flavour violating 
couplings in the mass eigenstate basis for quarks as a result of different 
transformation properties of the third generation under the extended gauge 
group. The flavour violating couplings of $Z^\prime$ are parametrized by 
 complex couplings $\Delta^{ij}_L(Z^\prime)$ with $i,j=d,s,b$ in the present 
paper.

When the small but non-vanishing mixing represented by $\sin\xi$ is 
taken into account, not only the flavour violating couplings of the mass eigenstate $Z_\mu^1$ to quarks are generated but also its flavour diagonal 
couplings to SM fermions differ from  the ones of the SM $Z$ 
boson. Explicitly we have for $i\not=j$
\be\label{ZZprime}
 \Delta^{ij}_L(Z^1)=\sin\xi \, \Delta^{ij}_L(Z^\prime)\equiv  \Delta^{ij}_L(Z), 
\qquad  \Delta^{ij}_L(Z^2)=\cos\xi \, \Delta^{ij}_L(Z^\prime)\approx\Delta^{ij}_L(Z^\prime),
\ee
where in order not to modify the notation in flavour violating observables 
relative to our previous papers we will still use 
$Z$ for $Z^1$ and $Z^\prime$ for $Z^2$ with masses $M_Z$ and $M_{Z^\prime}$, 
respectively. The small shifts in the masses of these gauge bosons relative 
to the case $\sin\xi=0$ are irrelevant in flavour violating processes. 

For flavour diagonal couplings to fermions { (generically denoted with $f$)} we have with $k=L,R,A,V$ 
\be\label{ffZ1}
\Delta^{ff}_k(Z^1)=\cos\xi \, \Delta^{ff}_k(Z)+\sin\xi \, \Delta^{ff}_k(Z^\prime),
\ee
\be\label{ffZ2}
\Delta^{ff}_k(Z^2)=\cos\xi \, \Delta^{ff}_k(Z^\prime)-\sin\xi \, \Delta^{ff}_k(Z)\,.
\ee 
In the calculations of flavour violating effects we can neglect the mixing 
effects in these couplings so that we can simply set 
\be\label{ffZ1a}
\Delta^{ff}_k(Z^1)=\Delta^{ff}_k(Z),\qquad
\Delta^{ff}_k(Z^2)=\Delta^{ff}_k(Z^\prime)
\ee
as in our previous papers, but in the discussion of electroweak precision 
tests in Section~\ref{EWP} we have to keep mixing effects in (\ref{ffZ1}). 
Following \cite{Altarelli:1996pr} in this case we will use for the modified diagonal  $Z$ couplings to fermions
\be\label{effZ}
[\Delta^{f}_k(Z)]_{\rm eff}\equiv\cos\xi \, \Delta^{ff}_k(Z)+\sin\xi \, \Delta^{ff}_k(Z^\prime).
\ee
The second term in this equation allows then as we will see in the course of 
our analysis to select by means of electroweak precision observables 
the favourite 331 models. 

Now the flavour violating $Z^\prime$ couplings to quarks, for the three meson systems  $K$, $B_d$ and $B_s$,
\be\label{Deltas}
\Delta^{sd}_L(Z'), \qquad \Delta^{bd}_L(Z'),  \qquad \Delta^{bs}_L(Z')\,
\ee
depend on the elements $v_{ij}$ of  a mixing matrix $V_L$.
Being proportional to $v_{32}^*v_{31}$, $v_{33}^*v_{31}$ and $v_{33}^*v_{32}$, respectively, they depend only on four new parameters (explicit formulae are given in \cite{Buras:2012dp}):
\be\label{newpar}
\tilde s_{13}, \quad \tilde s_{23}, \quad \delta_1, \quad \delta_2\,.
\ee
Here $\tilde s_{13}$ and $\tilde s_{23}$ are positive definite and $\delta_i$ in
the range $[0,2\pi]$.
Therefore for fixed  $M_{Z'}$ and $\beta$, the $Z'$ contributions to all processes
analyzed by us depend only on these parameters implying very strong
correlations between NP contributions to various observables.
Indeed, 
the $B_d$ system involves only the parameters ${\tilde s}_{13}$ and $\delta_1$ while the $B_s$ system depends  on
${\tilde s}_{23}$ and $\delta_2$. Moreover, stringent correlations between observables in $B_{d,s}$ sectors and in the kaon
sector are found since kaon physics depends on ${\tilde s}_{13}$, ${\tilde s}_{23}$ and $\delta_2 - \delta_1$.
A very constraining feature of this models is that the diagonal couplings
of $Z^\prime$ to quarks and leptons are fixed for a given $\beta$, except
for a weak dependence on $M_{Z^\prime}$ due to running of $\sin^2\theta_W$. 

As the mass $M_Z$ and flavour diagonal $Z$-couplings to all SM fermions are 
known, the model is also predictive after the inclusion of $Z-Z^\prime$ mixing, although one additional parameter, $\tan\bar\beta$, enters 
the game.
This mixing has been calculated in \cite{Diaz:2004fs}
in terms of the $SU(3)_L$ and $U(1)_X$  gauge couplings and the relevant vacuum 
expectations values but for our purposes it is useful to express it in terms of 
measurable quantities and $\beta$. Repeating this calculation 
we find an important expression
\be\label{sxi}
\sin\xi=\frac{c_W^2}{3} \sqrt{f(\beta)}\left(3\beta \frac{s_W^2}{c_W^2}+\sqrt{3}a\right)\left[\frac{M_Z^2}{M_{Z^\prime}^2}\right]
\equiv B(\beta,a) \left[\frac{M_Z^2}{M_{Z^\prime}^2}\right],
\ee
where 
\be\label{central}
f(\beta)=\frac{1}{1-(1+\beta^2)s_W^2} > 0,\,
\ee
$s_W^2=\sin^2\theta_W$ and 
\be\label{ratiov}
-1 < a=\frac{v_-^2}{v_+^2}< 1
\ee
 with $v^2_\pm$  given in terms of the 
vacuum expectation values of two  Higgs triplets $\rho$ and $\eta$ as follows
\be
v_+^2=v_\eta^2+v_\rho^2, \qquad v_-^2=v_\eta^2-v_\rho^2\,.
\ee

As the Higgs system responsible for the breakdown of the SM group has 
the structure of a two Higgs doublet model and the triplets $\rho$ and 
$\eta$ are responsible for the masses of up-quarks and down-quarks respectively 
one can express the parameter $a$ in terms of the usual $\tan\bar\beta$ where 
we introduced a {\it bar} to distinguish the usual angle $\beta$  from the 
parameter $\beta$ in 331 models. We have then 
\be\label{basica}
a=\frac{1-\tan^2\bar\beta}{1+\tan^2\bar\beta}, \qquad \tan\bar\beta=\frac{v_\rho}{v_\eta}.
\ee
Thus for $\tan\bar\beta=1$ the parameter $a=0$ which simplifies the 
formula for $\sin\xi$ relating uniquely its sign to the sign of $\beta$. 
On the other hand in the large $\tan\bar\beta$ limit we find $a=-1$ and 
in the low $\tan\bar\beta$ limit one has $a=1$.

 We have emphasized in  \cite{Buras:2013dea} that the couplings $\Delta^{ij}_L(Z^\prime)$ should be evaluated at $\mu=M_{Z^\prime}$ and 
this 
implies that the 
$s_W^2$ entering these couplings should be evaluated at $\mu=M_{Z^\prime}$. 
In evaluating  $\Delta^{ij}_L(Z)$ by means of (\ref{ZZprime}) such  $Z^\prime$-couplings 
should be used. However, as the $Z-Z^\prime$ mixing is generated in the 
process of the SM electroweak symmetry breaking, in evaluating $\sin\xi$ 
by means of (\ref{sxi}) and subsequently  $\Delta^{ij}_L(Z)$ by means of (\ref{ZZprime}) the value of $s_W^2$ at $\mu=M_Z$ should be used.

 Our result for $\sin\xi$ differs from the one that one would obtain from 
 the formula given in \cite{Diaz:2004fs,CarcamoHernandez:2005ka} by expressing it in terms of 
$s_W$, $c_W$, $M_Z$ and $M_{Z^\prime}$. We find opposite overall sign and 
the factor  $\sqrt{3}$ in front of the parameter $a$ that is missing 
in  \cite{Diaz:2004fs,CarcamoHernandez:2005ka}\footnote{The authors of these papers 
confirm our findings \cite{Martinez:2014lta}.}. The difference in sign is important for the 
interference between NP  contributions from $Z$ and $Z^\prime$ 
exchanges and consequently for the pattern of NP effects. It is also 
crucial for the interplay of flavour physics with electroweak precision tests 
 and should also
 have   an impact on the analyses of $Z-Z^\prime$ mixing 
effects in  \cite{Diaz:2004fs,CarcamoHernandez:2005ka,Ochoa:2005ih}. But these correlations 
depend also on the value of $\tan\bar\beta$ and we will see  this explicitly 
below.

The expression in (\ref{sxi}) tells us indeed that $\sin\xi$ is very small but one should remember that the propagator suppression of FCNC transitions in the case of $Z^\prime$ is by a factor of $M_{Z^\prime}^2/M_Z^2$ stronger than in the case of $Z$ at the amplitude level. Therefore we should make a closer look at the values of $\sin\xi$ and $Z^\prime$ couplings to leptons as functions of 
$\beta$ and $\tan\bar\beta$ and compare them with the known $Z$ couplings to fermions in order to decide whether $Z$ boson contributions to FCNC processes can be neglected or not. 
However first we have to elaborate on the choice of fermion representations.

\boldmath
\subsection{Choice of Fermion Representations}\label{Frepresentation}
\unboldmath
As already emphasized in \cite{Diaz:2004fs,CarcamoHernandez:2005ka} the choice of $\beta$ 
does not uniquely specify the phenomenology of the 331 model considered 
which further depends on the choice of fermion representations under 
$SU(3)_L$. Here we discuss some aspects of this dependence that are relevant 
for our study.

Our choice of representations in \cite{Buras:2012dp,Buras:2013dea} under 
$SU(3)_L$ can be summarized as follows. The first two generations of 
quarks are put into triplets ($3$) while the third one into the antitriplet 
$(3^*)$:
\bea
\left(\begin{array}{c}
u  \\
 d   \\
D  \\
\end{array}\right)_L \hskip 2 cm
\left(\begin{array}{c}
c  \\
 s   \\
S  \\
\end{array}\right)_L \hskip 2 cm
\left(\begin{array}{c}
b  \\
 -t   \\
T  \\
\end{array}\right)_L .\hskip 2 cm
\label{quarksL}
\eea
 The corresponding right handed quarks are singlets. The anomaly 
cancellation then requires that leptons are put into antitriplets:
\bea
\left(\begin{array}{c}
e  \\
 -\nu_e   \\
E_e  \\
\end{array}\right)_L \hskip 2 cm
\left(\begin{array}{c}
\mu  \\
 -\nu_\mu   \\
E_\mu  \\
\end{array}\right)_L \hskip 2 cm
\left(\begin{array}{c}
\tau  \\
 -\nu_\tau   \\
E_\tau  \\
\end{array}\right)_L .\hskip 2 cm
\,\, \eea 
We refer to this choice as $F_1$.

On the other hand in \cite{Diaz:2004fs,CarcamoHernandez:2005ka} the 
 triplets and antitriplets are interchanged relative to our choice. 
That is the first two quark generations are in antitriplets while the third 
one in a triplet. Therefore leptons are also in triplets. We call this 
fermion assignment $F_2$\footnote{In  \cite{Diaz:2004fs,CarcamoHernandez:2005ka} still 
two other quark assignments are discussed in which the first or the second 
quark generation transforms differently under $SU(3)_L$ than the remaining 
two. But we find the ones listed above more natural due to large top quark 
mass and we do not discuss these two additional possibilities.}.

The important two features to be remembered for our discussion below is that 
for a given $\beta$:
\begin{itemize}
\item
The expression for $\sin\xi$ in (\ref{sxi}) is independent  of whether $F_1$ or 
$F_2$ is used.
\item 
On the other hand as evident from the comparison of our compendium for 
$Z^\prime$ couplings to fermions in \cite{Buras:2013dea}
with Table~4 of  \cite{CarcamoHernandez:2005ka}
the signs in front of $\beta$ in these couplings are changed when going 
from $F_1$ to $F_2$.  
This property can be derived from the action of the 
relevant operator $\hat Q_W$ on triplet and antitriplet. See formulae in 
Section 2 of \cite{Buras:2012dp}.
\end{itemize}

These observations have the following important phenomenological implications 
given here first without FCNCs due to $Z$ boson:
\begin{itemize}
\item
In $F_1$ scenario the models with  $\beta=-2/\sqrt{3}$ and $\beta=-1/\sqrt{3}$ are useful for the explanation of the anomalies in $B_d\to K^*\mu^+\mu^-$ because with $F_1$ representations the coupling $\Delta_V^{\mu\bar\mu}(Z^\prime)$  is large. On the other hand the models with $\beta=2/\sqrt{3}$ and $\beta=1/\sqrt{3}$ having significant $\Delta_A^{\mu\bar\mu}(Z^\prime)$ coupling provide interesting 
NP effects in  $B_{s,d}\to\mu^+\mu^-$.
\item
In $F_2$ scenario the situation is reversed. The models with 
$\beta=2/\sqrt{3}$ and $\beta=1/\sqrt{3}$ are useful for  the explanation of the anomalies in $B_d\to K^*\mu^+\mu^-$ while the ones with 
$\beta=-2/\sqrt{3}$ and $\beta=-1/\sqrt{3}$ for  $B_{s,d}\to\mu^+\mu^-$.
\item
While these two scenarios cannot be distinguished by flavour observables 
when only $Z^\prime$ contributions are considered they can be distinguished 
when $Z$ boson contributions are taken into account. This originates in 
the fact that  the $\sin\xi$ entering the 
$\Delta_L^{ij}(Z)$ couplings in (\ref{ZZprime}) {\it does depend} on the 
sign of $\beta$ but {\it does not depend} on whether $F_1$ scenario or $F_2$ 
scenario is considered.  In other words the invariance in flavour observables under the transformations
\be\label{symmetry}
\beta \rightarrow -\beta, \qquad F_1 \rightarrow F_2
\ee
present in the absence of $Z-Z^\prime$ mixing is broken by this mixing. 
We will see this explicitly in our numerical analysis below.
\item
As a particular sign of $\beta$ could 
be favoured by flavour conserving observables, in particular electroweak 
precision tests, this feature allows in principle to determine  whether 
the representation $F_1$ or $F_2$ is favoured by nature. We will see this 
explicitly in Section~\ref{EWP}.
\end{itemize}

\boldmath
\subsection{$\Delta F=2$ Processes}
\unboldmath
In the models considered only SM $\Delta F=2$ operator  { (i.e. that change flavour quantum number by two units, as for example in neutral meson mixing)} in each meson system  is present and the effects of NP in 
all $\Delta F=2$ transitions can be compactly summarized by generally 
flavour dependent 
shifts $\Delta S$ in the SM one loop function $S$ that is flavour independent. 
However due to the relation (\ref{ZZprime}) the flavour dependence of 
the shifts  $\Delta S$ due to $Z$ and $Z^\prime$ contributions is the same. Consequently 
for all meson systems the ratio of the shifts in $S$ due to $Z$ and $Z^\prime$ 
is given universally  as follows:
\be\label{DF2}
\frac{\Delta S(Z)}{\Delta S(Z^\prime)}=\sin^2\xi \left[\frac{M_{Z\prime}^2}{M_{Z}^2}\right]=B^2(\beta,a)\left[\frac{M_Z^2}{M_{Z^\prime}^2}\right]. 
\ee
As $B(\beta,a)\le 1.1$ in all four  
models considered by us, it follows 
that $Z$ contributions to all $\Delta F=2$ transitions can be neglected. 
This is good news: the determination of the allowed values of the new 
parameters (\ref{newpar}) by means of $\Delta F=2$ processes remains 
unmodified relative to our analyses in \cite{Buras:2012dp,Buras:2013dea}. 
\boldmath
\subsection{$\Delta F=1$ Processes}
\unboldmath

It should be noted that in $\Delta F=2$ processes the flavour violating coupling of $Z$ enters twice 
which resulted in $\sin^2\xi$ dependence in the $\Delta F=2$ amplitudes. However, 
in $\Delta F=1$ amplitudes { (implying a change by one unit of flavour quantum number, as in weak decays)} it appears only once, whereas the dependence on the mass 
of the exchanged gauge boson remains unchanged. Again the flavour dependence in the 
vertex involving quarks is the same for $Z$ and $Z^\prime$ and as the operators 
in each systems are also the same, the ratio of the amplitudes $A_{\ell\ell}(Z)$
and $A_{\ell\ell}(Z^\prime)$ takes a very simple form:
\be\label{DF1}
R^k_{\ell\ell}=\frac{A_{\ell\ell}(Z)}{A_{\ell\ell}(Z^\prime)}=\sin\xi \left[\frac{M_{Z\prime}^2}{M_{Z}^2}\right] \left[\frac{\Delta^{\ell\ell}_k(Z)}{\Delta^{\ell\ell}_k(Z^\prime)}\right]= B(\beta,a)\left[\frac{\Delta^{\ell\ell}_k(Z)}{\Delta^{\ell\ell}_k(Z^\prime)}\right],
\ee
where $k=L,R,A,V$ and $\ell\ell$ stands either for charged leptons or 
neutrinos in the final state.
The remarkable property of this formula is its independence on $M_{Z^\prime}$. 
Consequently the ratios $R^k_{\ell\ell}$ with known $Z$ couplings to leptons are only  functions of $\beta$ and of the parameter $a$ or 
equivalently $\tan\bar\beta$. In addition they depend on whether 
the representation $F_1$ or $F_2$ is considered.

 The ratios $R^k_{\ell\ell}$ give us the information on the importance of $Z$ 
contributions relatively to $Z^\prime$ contributions but in order to get 
the full picture, in particular in view of the dependence of NP effects 
on the choice of fermion { representations,} it is useful to consider the 
quantities
\be\label{WF1}
W^k_{\ell\ell}=\Delta^{\ell\ell}_k(Z^\prime)\left(1+R^k_{\ell\ell}\right),
\ee
which will directly enter the phenomenological expressions.

In Table~\ref{tab:Rmumu} we show the values of $\sin\xi$, 
$R^k_{\ell\ell}$  and $W^k_{\ell\ell}$  relevant for the couplings $\Delta_V^{\mu\mu}$, $\Delta_A^{\mu\mu}$ 
and $\Delta_L^{\nu\nu}$ in scenario $F_1$ for fermion 
representations 
for the four values of $\beta$ and two values of $\tan\bar\beta=1(5)$. 
The corresponding results for scenario $F_2$ are
given in Table~\ref{tab:Rmumu2}  and for $\tan\bar\beta=0.2$ and $F_1(F_2)$ 
in Table~\ref{tab:Rmumu3}.
{ In these tables we fix $M_{Z^\prime}=3$ TeV, as we do in our numerical analysis.}

\begin{figure}[!tb]
\centering
\includegraphics[width = 0.5\textwidth]{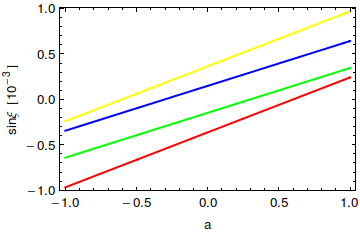}
\caption{\it $\sin\xi$ as a function of $a$ for different $\beta$ ($-\frac{2}{\sqrt{3}}$: red, $-\frac{1}{\sqrt{3}}$: green, 
$\frac{1}{\sqrt{3}}$: blue,  $\frac{2}{\sqrt{3}}$: yellow)  {and for $M_{Z^\prime}=3$ TeV}.\label{fig:six}} 
\end{figure}

 In Fig.~\ref{fig:six} we show $\sin\xi$  as a function of $a$ for 
different values of $\beta$. The values $a=0$ and $a=-12/13(12/13)$ correspond 
to $\tan\bar\beta=1$ and $\tan\bar\beta=5(0.2)$, respectively.

\begin{table}[!tb]
{\renewcommand{\arraystretch}{1.3}
\begin{center}
\begin{tabular}{|c||c|c|c|c|}
\hline
$\beta$ & $-\frac{2}{\sqrt{3}}$ & $-\frac{1}{\sqrt{3}}$ & $\frac{1}{\sqrt{3}}$ & $\frac{2}{\sqrt{3}}$ \\
\hline
$\sin\xi\,\,[10^{-3}]$ & $-0.36(-0.92)$ & $-0.15(-0.60)$ & $0.15(-0.31)$& $0.36(-0.19)$\\ 
$R_V^{\mu\mu}$ & $0.015(0.038)$ & $0.012(0.047)$ & $-4.36(9.02)$ & $0.046(0.024)$ \\
$R_A^{\mu\mu}$ & $1.77(4.50)$ & $0.46(1.88)$ & $-0.23(0.48)$ & $-0.36(0.19)$\\
$R_L^{\nu\nu}$ & $-0.36(-0.91)$ & $-0.23(-0.94)$ & $0.46(-0.95)$ & $1.77(-0.95)$ \\
$W_V^{\mu\mu}$ & $0.74(0.76)$ & $0.39(0.41)$ & $-0.0035(0.010)$ & $-0.25(-0.24)$ \\
$\Delta_V^{\mu \mu}(Z^\prime)$ &  0.731 & 0.386 & 0.001 & $-0.242$ \\
$W_A^{\mu\mu}$ & $-0.23(-0.45)$ & $-0.19(-0.37)$ & $-0.20(-0.38)$ & $-0.26(-0.49)$\\
$\Delta_A^{\mu \mu}(Z^\prime)$ & $-0.082$ &$-0.130$ & $-0.258$ & $-0.407$ \\
$W_L^{\nu\nu}$ & $0.26(0.036)$ & $0.20(0.015)$ & $0.19(0.0060)$ & $0.23(0.0042)$\\
$\Delta_L^{\nu\nu}(Z^\prime)$ & 0.407 &0.258 & 0.130 & 0.082 \\
$R_{\varepsilon^\prime}$ & $-0.12(-0.31)$ &$-0.12(-0.50)$ &$-0.12(0.25)$ &$-0.12(0.066)$  \\
\hline
\end{tabular}
\end{center}}
\caption{\it $\sin\xi$, $R_{V,A,L}^{\mu\mu}$, $W_{V,A,L}^{\mu\mu}$, $\Delta_{V,A,L}^{\mu\mu,\nu\nu}$ and $R_{\varepsilon^\prime}$ from (\ref{equ:Repsprime}) 
for different $\beta$ and $\tan\bar\beta=1(5)$ in scenario $F_1$ for fermion representations { and for $M_{Z^\prime}=3$ TeV}. $R_{\varepsilon^\prime}$ is 
defined in  (\ref{equ:Repsdef}).
\label{tab:Rmumu}}~\\[-2mm]\hrule
\end{table}

\begin{table}[!tb]
{\renewcommand{\arraystretch}{1.3}
\begin{center}
\begin{tabular}{|c||c|c|c|c|}
\hline
$\beta$ & $-\frac{2}{\sqrt{3}}$ & $-\frac{1}{\sqrt{3}}$ & $\frac{1}{\sqrt{3}}$ & $\frac{2}{\sqrt{3}}$ \\
\hline
$\sin\xi\,\,[10^{-3}]$ & $-0.36(-0.92) $ & $-0.15(-0.60) $ & $0.15(-0.31) $& $0.36(-0.19) $\\ 
$R_V^{\mu\mu}$ & $-0.046(-0.12) $ & $4.36(17.7) $ & $-0.012(0.024) $ & $-0.015(0.0081) $ \\
$R_A^{\mu\mu}$ & $ 0.36(0.91)$ & $0.23(0.94) $ & $-0.46(0.95) $ & $-1.77(0.95) $\\
$R_L^{\nu\nu}$ & $-1.77(-4.50) $ & $-0.46(-1.87) $ & $0.23(-0.48) $ & $0.36(-0.19) $ \\
$W_V^{\mu\mu}$ & $-0.23(-0.21) $ & $ 0.0055(0.019)$ & $0.38(0.40) $ & $ 0.72(0.74)$ \\
$\Delta_V^{\mu \mu}(Z^\prime)$ & $-0.242$  & 0.001 &  0.386 & 0.731  \\
$W_A^{\mu\mu}$ & $-0.55(0.78) $ & $-0.32(-0.50) $ & $-0.070(-0.25) $ & $ 0.064(-0.16)$\\
$\Delta_A^{\mu \mu}(Z^\prime)$ & $-0.407$ & $-0.258$ & $-0.130$ & $-0.082$ \\
$W_L^{\nu\nu}$ & $-0.064(-0.29) $ & $0.070(-0.11) $ & $0.32(0.13) $ & $ 0.55(0.33)$\\
$\Delta_L^{\nu\nu}(Z^\prime)$ &$0.082$ &$0.130$ & $0.258$ &  $0.407$ \\
$R_{\varepsilon^\prime}$ & $0.12(0.31)$ &$0.12(0.50)$ &$0.12(-0.25)$ &$0.12(-0.066)$  \\
\hline
\end{tabular}
\end{center}}
\caption{\it $\sin\xi$, $R_{V,A,L}^{\mu\mu}$, $W_{V,A,L}^{\mu\mu}$, $\Delta_{V,A,L}^{\mu\mu,\nu\nu}$  and $R_{\varepsilon^\prime}$ from (\ref{equ:Repsprime})
 for different $\beta$ and $\tan\bar\beta=1(5)$ in scenario $F_2$ for fermion representations  { and for $M_{Z^\prime}=3$ TeV}. $R_{\varepsilon^\prime}$ is defined in  (\ref{equ:Repsdef}).
\label{tab:Rmumu2}}~\\[-2mm]\hrule
\end{table}
\begin{table}[!tb]
{\renewcommand{\arraystretch}{1.3}
\begin{center}
\begin{tabular}{|c||c|c|c|c|}
\hline
$\beta$ & $-\frac{2}{\sqrt{3}}$ & $-\frac{1}{\sqrt{3}}$ & $\frac{1}{\sqrt{3}}$ & $\frac{2}{\sqrt{3}}$ \\
\hline
$\sin\xi\,\,[10^{-3}]$ & $0.194 (0.194)$ & $0.307 (0.307)$ & $0.603 (0.603)$& $0.921 (0.921)$\\ 
$R_V^{\mu\mu}$ & $-0.008 (0.024)$ & $-0.024 (-9.02)$ & $-17.73 (-0.047)$ & $0.115 (-0.038)$ \\
$R_A^{\mu\mu}$ & $-0.950 (-0.192)$ & $-0.954 (-0.479)$ & $-0.942 (-1.876)$ & $-0.912 (-4.500)$\\
$R_L^{\nu\nu}$ & $0.192 (0.950)$ & $0.479 (0.954)$ & $1.876 (0.942)$ & $4.500 (0.912)$ \\
$W_V^{\mu\mu}$ & $0.725 (-0.248)$ & $0.377 (-0.008)$ & $-0.017 (0.368)$ & $-0.270 (0.703)$ \\
$W_A^{\mu\mu}$ & $-0.004 (-0.328)$ & $-0.006 (-0.134)$ & $-0.015 (0.113)$ & $-0.036 (0.288)$\\
$W_L^{\nu\nu}$ & $0.485 (0.160)$ & $0.381 (0.253)$ & $0.372 (0.501) $ & $ 0.453(0.778)$\\
$R_{\varepsilon^\prime}$ & $0.066(-0.066)$ &$0.25(-0.25)$ &$-0.50(0.50)$ &$-0.31(0.31)$  \\
\hline
\end{tabular}
\end{center}}
\caption{\it $\sin\xi$, $R_{V,A,L}^{\mu\mu}$, $W_{V,A,L}^{\mu\mu}$ and $R_{\varepsilon^\prime}$ from Eq.~(\ref{equ:Repsprime}) for different $\beta$ and $\tan\bar\beta=0.2$ in scenario $F_1$ ($F_2$) for fermion 
representations { and for $M_{Z^\prime}=3$ TeV}. $R_{\varepsilon^\prime}$ is  defined in  (\ref{equ:Repsdef}).
\label{tab:Rmumu3}}~\\[-2mm]\hrule
\end{table}

We observe the following features:
\begin{itemize}
\item
In the case of  decays involving the coupling $\Delta_V^{\mu\mu}(Z)$ the contributions from $Z$ boson can be neglected due 
its  small vectorial coupling to charged leptons. The large values of $R^V_{\mu\mu}$  for $\beta=1/\sqrt{3}$ { in the case of $F_1$ and  
$\beta=-1/\sqrt{3}$ in the case of $F_2$  do not} imply large contribution of $Z$ boson as 
in this case $Z^\prime$ contribution is negligible. This is good news. 
The explanation of $B_d\to K^*\mu^+\mu^-$ anomalies with $Z^\prime$ contributions  presented in  \cite{Buras:2013dea} remains basically unmodified.
\item
But in the case of 
 $B_{s,d}\to \mu^+\mu^-$, $B_d\to K\mu^+\mu^-$ and decays
 with neutrinos in the final state, like  $b\to s\nu\bar\nu$ transitions, $\kpn$ and $\klpn$  $Z$ contributions cannot be generally neglected but 
the size of the additional contributions depends on $\beta$ and $\tan\bar\beta$.
\item
Comparing Tables~\ref{tab:Rmumu} and \ref{tab:Rmumu2} we observe that 
indeed the symmetry (\ref{symmetry}) is broken by $Z-Z^\prime$ mixing. 
This is also seen in  Table~\ref{tab:Rmumu3}.
\item
Finally, we emphasize that the pattern of NP effects 
is governed by the sign of $\sin\xi$ in (\ref{sxi}).
\end{itemize}
In the next section we will investigate these new contributions numerically.

\boldmath
\section{$Z$ Contributions to $\Delta F=1$ Observables}\label{sec:3}
\unboldmath
\subsection{Preliminaries}
The inclusion of flavour violating effects from $Z$ in the observables analyzed in \cite{Buras:2013dea} amounts to replacing  the 
 shifts due to NP in SM one-loop functions, given in the formulae (22)-(27) in that paper, by the following ones. 

Defining 
\be\label{gsm}
g_{\text{SM}}^2=4\frac{G_F}{\sqrt 2}\frac{\alpha}{2\pi\sin^2\theta_W}\,, \qquad
\lambda^{(K)}_i=V_{is}^*V_{id},\qquad \lambda_t^{(q)}=V_{tb}^*V_{tq}
\ee
one has
for decays $B_q\to\mu^+\mu^-$ with $q=d,s$  governed by the function $Y$ 
\be\label{DYB}
\Delta Y(B_q)=
\left[\frac{\Delta_{A}^{\mu\bar\mu}(Z')}{M_{Z'}^2g^2_{\rm SM}}\right]
\frac{\Delta_{L}^{qb}(Z')}{ V_{tq}^\ast V_{tb}}(1+R^A_{\mu\mu})
\ee
and for $K_L\to\mu^+\mu^-$
\be\label{DYK}
\Delta Y(K)=
\left[\frac{\Delta_{A}^{\mu\bar\mu}(Z')}{M_{Z'}^2g^2_{\rm SM}}\right]
\frac{\Delta_{L}^{sd}(Z')}{ V_{ts}^\ast V_{td}} (1+R^A_{\mu\mu}).
\ee

Similarly for $b\to q\nu\bar\nu$ transitions  governed by the function $X$ one finds
\be\label{DXB}
\Delta X(B_q)=
\left[\frac{\Delta_{L}^{\nu\nu}(Z')}{g^2_{\rm SM}M_{Z'}^2}\right]
\frac{\Delta_{L}^{qb}(Z')}{ V_{tq}^\ast V_{tb}}(1+R^L_{\nu\nu})
\ee
and for $\kpn$ and $\klpn$
\be\label{XLK}
\Delta X(K)=\left[\frac{\Delta_L^{\nu\bar\nu}(Z')}{g^2_{\rm SM}M_{Z'}^2}\right]
    \frac{\Delta_L^{sd}(Z')}{V_{ts}^* V_{td}} (1+R^L_{\nu\nu}).
\ee

The corrections from NP to the Wilson coefficients $C_9$ and $C_{10}$ {that weight the semileptonic operators in the effective hamiltonian} relevant
for $b\to s\mu^+\mu^-$ transitions and used in the recent literature
are given as follows
\begin{align}
 \sin^2\theta_W C^{\rm NP}_9 &=-\frac{1}{g_{\text{SM}}^2M_{Z^\prime}^2}
\frac{\Delta_L^{sb}(Z')\Delta_V^{\mu\bar\mu}(Z')} {V_{ts}^* V_{tb}}(1+R^V_{\mu\mu}) ,\label{C9}\\
   \sin^2\theta_W C^{\rm NP}_{10} &= -\frac{1}{g_{\text{SM}}^2M_{Z^\prime}^2}
\frac{\Delta_L^{sb}(Z')\Delta_A^{\mu\bar\mu}(Z')}{V_{ts}^* V_{tb}}(1+R^A_{\mu\mu})\label{C10}.
 \end{align}
{ As seen in these equations $C^{\rm NP}_9$ involves leptonic vector coupling 
of $Z^\prime$ while $C^{\rm NP}_{10}$ the axial-vector one. $C^{\rm NP}_9$ is 
crucial for $B_d\to K^*\mu^+\mu^-$, $C^{\rm NP}_{10}$ for $B_s\to\mu^+\mu⁻$ and 
both coefficients are relevant for $B_d\to K\mu^+\mu^-$.}

\boldmath
\subsection{Numerical Results for $B$ Decays}
\unboldmath

\begin{figure}[!tb]
 \centering
\includegraphics[width = 0.44\textwidth]{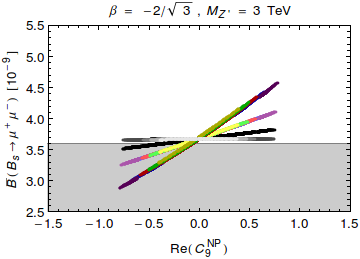}
\includegraphics[width = 0.44\textwidth]{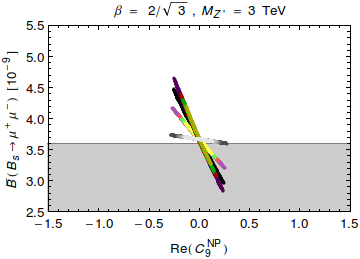}
\\
\includegraphics[width = 0.44\textwidth]{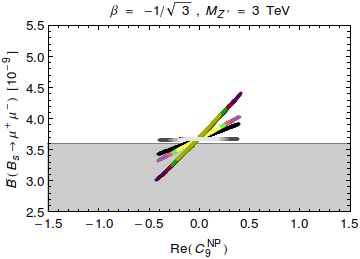}
\includegraphics[width = 0.44\textwidth]{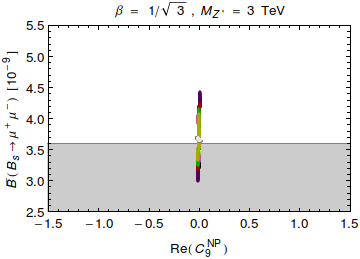}
\caption{\it Correlation
 $\overline{\mathcal{B}}(B_s\to\mu^+\mu^-)$ versus $\RE(C_9^\text{NP})$ in models with $F_1$ for
$\beta=\pm 1/\sqrt{3}$ and $\beta=\pm 2/\sqrt{3}$ setting $M_{Z^\prime}=3\tev$ and different
values of $C_{B_s}$  ($C_{B_s}=0.90\pm0.01, \, 0.96\pm 0.01,\, 1.00\pm 0.01, \,
C_{B_s}= 1.04\pm 0.01, \,1.10\pm 0.01$ (yellow, green, red, blue, purple; from light gray to dark gray)). 
Black is without $Z-Z^\prime$-mixing, lighter colours are for 
$\tan\bar\beta = 1$, darker colours for $\tan\bar\beta = 5$ and gray colours for $\tan\bar\beta=0.2$. The gray regions show  the 
experimental range  
$\overline{\mathcal{B}}(B_{s}\to\mu^+\mu^-) = (2.9\pm0.7)\cdot 10^{-9}$.
}\label{fig:ReC9vsBsmu}~\\[-2mm]\hrule
\end{figure}

\begin{figure}[!tb]
 \centering
\includegraphics[width = 0.44\textwidth]{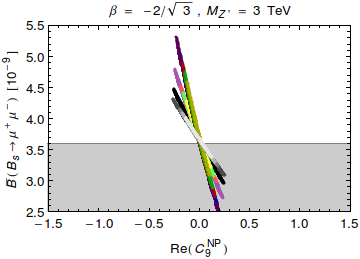}
\includegraphics[width = 0.44\textwidth]{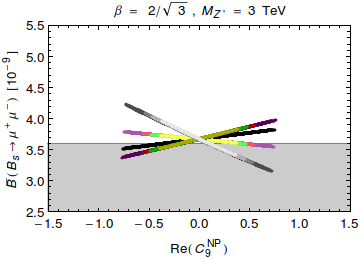}
\\
\includegraphics[width = 0.44\textwidth]{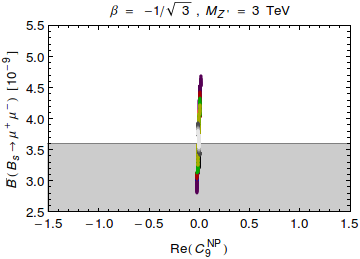}
\includegraphics[width = 0.44\textwidth]{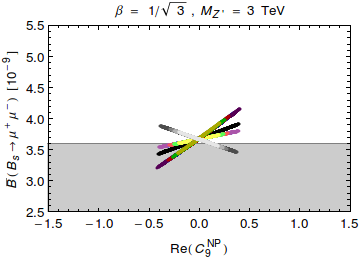}
\caption{\it As in Fig.~\ref{fig:ReC9vsBsmu} but for F2.
}\label{fig:ReC9vsBsmuF2}~\\[-2mm]\hrule
\end{figure}

\begin{figure}[!tb]
 \centering
\includegraphics[width = 0.45\textwidth]{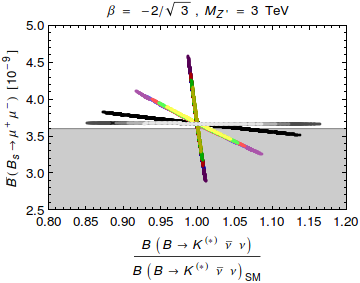}
\includegraphics[width = 0.45\textwidth]{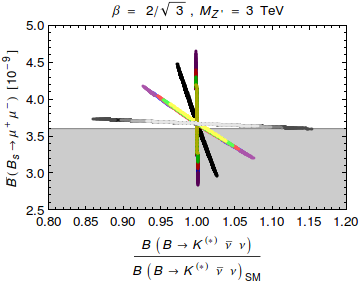}

\includegraphics[width = 0.45\textwidth]{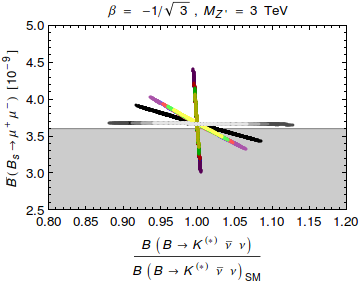}
\includegraphics[width = 0.45\textwidth]{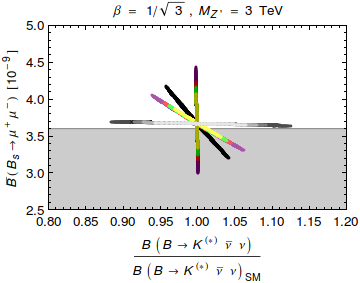}
\caption{ \it $\overline{\mathcal{B}}(B_s\to\mu^+\mu^-)$ versus the ratio ${\mathcal{B}(B\to K \nu \bar \nu)}/{ \mathcal{B}(B\to K \nu \bar 
\nu)_{\rm SM}}$  for all four $\beta=\pm\frac{2}{\sqrt{3}},
\pm
\frac{1}{\sqrt{3}}$. Colours as in Fig.~\ref{fig:ReC9vsBsmu}
}\label{fig:BsmuRnu}~\\[-2mm]\hrule
\end{figure}

\begin{figure}[!tb]
 \centering
\includegraphics[width = 0.45\textwidth]{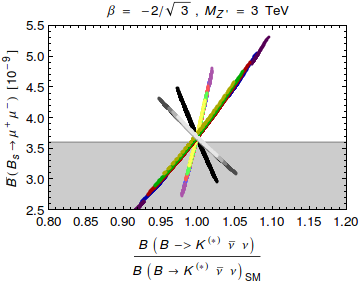}
\includegraphics[width = 0.45\textwidth]{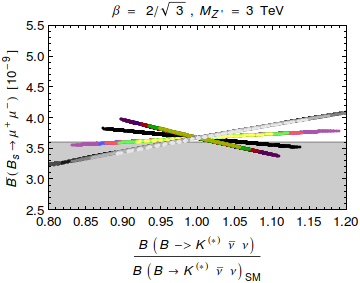}

\includegraphics[width = 0.45\textwidth]{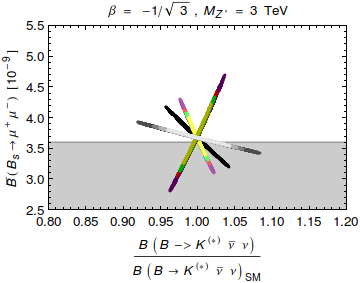}
\includegraphics[width = 0.45\textwidth]{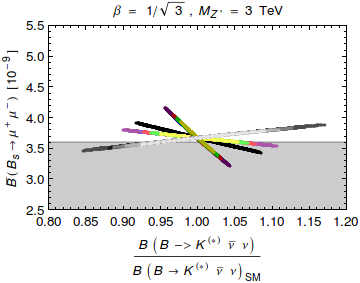}
\caption{ \it As in Fig.~\ref{fig:BsmuRnu} but for F2. 
}\label{fig:BsmuRnuF2}~\\[-2mm]\hrule
\end{figure}

\begin{figure}[!tb]
 \centering
\includegraphics[width = 0.45\textwidth]{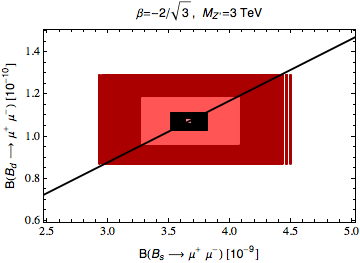}
\includegraphics[width = 0.45\textwidth]{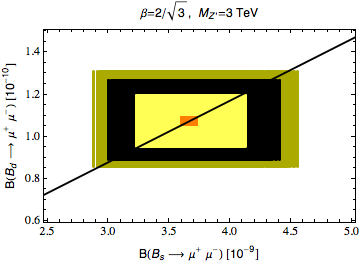}

\includegraphics[width = 0.45\textwidth]{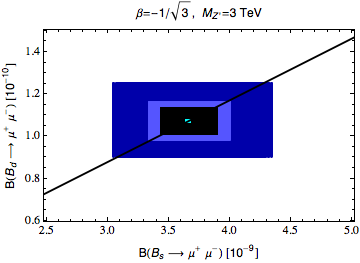}
\includegraphics[width = 0.45\textwidth]{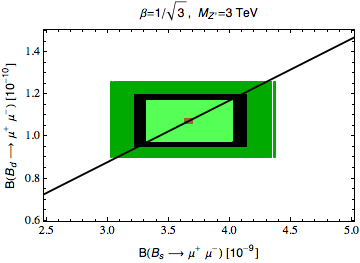}
\caption{\it Correlation
${\mathcal{B}}(B_d\to\mu^+\mu^-)$  versus
$\overline{\mathcal{B}}(B_s\to\mu^+\mu^-)$, without $Z-Z^\prime$ mixing (black), $\tan\bar\beta = 1$ (lighter colours), $\tan\bar\beta = 
5$ (darker colours)  and $\tan\bar\beta$ = 0.2 (small coloured box in the middle). The black line corresponds to the correlation in CMFV.
}\label{fig:BdvsBs}~\\[-2mm]\hrule
\end{figure}

\begin{figure}[!tb]
 \centering
\includegraphics[width = 0.45\textwidth]{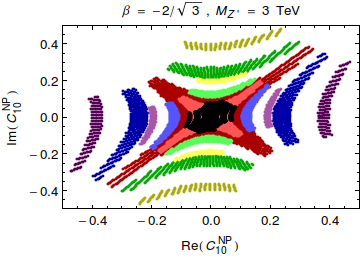}
\includegraphics[width = 0.45\textwidth]{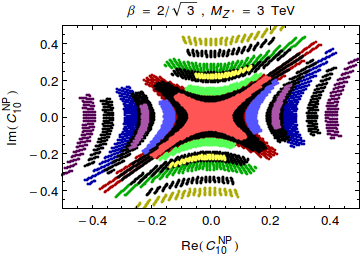}
\includegraphics[width = 0.45\textwidth]{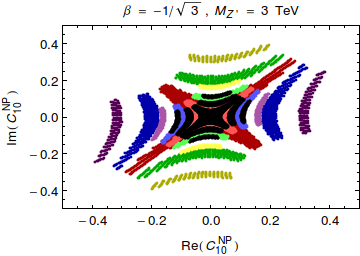}
\includegraphics[width = 0.45\textwidth]{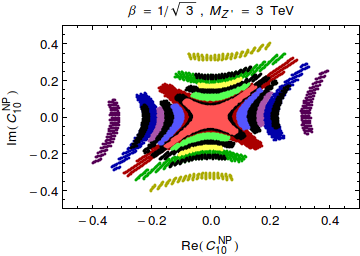}
\caption{ \it $\IM(C_{10}^{\rm NP})$ versus $\RE(C_{10}^{\rm NP})$
for $\beta=\pm 1/\sqrt{3}$ and
$\beta=\pm 2/\sqrt{3}$ setting $M_{Z^\prime}=3\tev$. Colour coding as in Fig.~\ref{fig:ReC9vsBsmu} but without $\tan\bar\beta = 0.2$ for which NP effects are very small.
}\label{fig:c10}~\\[-2mm]\hrule
\end{figure}

Before presenting our results we recall the present SM value for 
$B_s\to\mu^+\mu^-$
\cite{Bobeth:2013uxa}
\be\label{LHCb2}
 \overline{\mathcal{B}}(B_{s}\to\mu^+\mu^-)_{\rm SM}=(3.65\pm0.23)\times 10^{-9},\quad
\overline{\mathcal{B}}(B_{s}\to\mu^+\mu^-)_{\rm exp} = (2.9\pm0.7) \times 10^{-9}, \ee
where we have also shown the latest average of the results from LHCb and CMS
\cite{Aaij:2013aka,Chatrchyan:2013bka,CMS-PAS-BPH-13-007}. 
The agreement of  the SM prediction with the data for $B_s\to\mu^+\mu^-$
in (\ref{LHCb2}) is remarkable, although the rather large 
experimental error still allows for sizable NP contributions with the ones 
suppressing the branching ratio relative to its SM value being 
favoured. 

As far as the anomalies in $B_d\to K^*\mu^+\mu^-$  \cite{Aaij:2013iag,Aaij:2013qta} are concerned a number of analyses, of which we only quote three \cite{Descotes-Genon:2013wba,Altmannshofer:2013foa,Beaujean:2013soa}, indicate 
that $\RE(C_9^\text{NP})=-1.0\pm0.5$. Thus in the plots presented below the 
results with
\be
 \overline{\mathcal{B}}(B_{s}\to\mu^+\mu^-)< \overline{\mathcal{B}}(B_{s}\to\mu^+\mu^-)_{\rm SM},\qquad \RE(C_9^\text{NP})\le -0.5
\ee
are favoured by the present data.

The results for various observables in 331 models with fermion representations 
$F_1$ have been presented in Figs. 5-17 in \cite{Buras:2013dea}. The present 
analysis shows how the latter results are modified when $Z$ boson contributions 
are included and when the fermion representations $F_2$ are considered instead 
of $F_1$. 

First of all taking into account that the $Z$ 
contributions can be neglected in all $\Delta F=2$ transitions and in 
the coefficient $C_9$ the results in Figs.~7 and  8 in  
\cite{Buras:2013dea} remain basically unchanged as they involve only 
$\Delta F=2$ observables and $C_9$. 
 On the other hand 
in the processes in which NP is governed 
by the shifts in (\ref{DYB})--(\ref{XLK}) and $C_{10}$ we find 
that modifications can be sizable, in particular when two observables 
taking part in the correlation are affected by $Z$ contributions in 
a rather different manner.

In order to have appropriate comparison with the results in \cite{Buras:2013dea}  we use the same treatment of CKM parameters and hadronic uncertainties as 
in the latter paper so that the difference between various correlations 
are only due to differences in NP contributions. For this reason we do not 
list the input parameters that can be found in Table~3 of that paper.

The colour coding in the 
plots presented in this subsection is as follows:
\begin{itemize}
\item
The results of \cite{Buras:2013dea} and those new ones in fermion representations $F_2$ that include only $Z^\prime$ contributions are presented in {\it black}.
\item
The results that include both $Z^\prime$ and $Z$ contributions are given 
in colours that distinguish between the values of $C_{B_s} = \Delta M_s/(\Delta M_s)_\text{SM}$. 
\item
As for a given $\beta$ the contributions from $Z$ boson depend on $\tan\bar\beta$, we show the results for $\tan\bar\beta=1$ in light colours, for 
 $\tan\bar\beta=5$ in darker colours and for   $\tan\bar\beta=0.2$ in gray  
colours.
\item
Finally we show the results for the four different values $\beta$ in question 
and the fermion representations $F_1$ and $F_2$.
\end{itemize}

The results of this extensive numerical analysis are shown in Figs.~\ref{fig:ReC9vsBsmu}--\ref{fig:BdvsBs}. While with the comments just made these figures 
are self-explanatory, we would like to emphasize the most interesting 
features in them:
\begin{itemize}
\item
Comparison of Figs.~\ref{fig:ReC9vsBsmu} and \ref{fig:ReC9vsBsmuF2} 
demonstrates the breakdown of the invariance under (\ref{symmetry}) 
by $Z-Z^\prime$ mixing. 
As seen in Figs.~\ref{fig:BsmuRnu} and  \ref{fig:BsmuRnuF2}
this breakdown is even larger when channels with neutrinos in the final 
state are considered.
\item 
From the present perspective,  ignoring at first the constraints from electroweak precision observables, the most interesting model is the one with 
$\beta=-2/\sqrt{3}$ and fermion representations $F_1$ considered also by 
us in \cite{Buras:2013dea}. It allows to bring the theory closer to the 
data on $B_d\to K^*\mu^+\mu^-$ and $B_s\to\mu^+\mu^-$ than it is possible 
in the remaining models. In particular the inclusion of $Z$ boson contributions 
allows to suppress $\overline{\mathcal{B}}(B_s\to\mu^+\mu^-)$ by $(15-20)\%$ 
below its SM value, which is not possible if only $Z^\prime$ contributions 
are present. But this suppression is only significant for $\tan\bar\beta>1.0$ 
and is clearly visible for $\tan\bar\beta=5.0$. On the other hand for 
$\tan\bar\beta=0.2$ there is a destructive interference between $Z$ and $Z^\prime$ so that in this case NP effects in $\overline{\mathcal{B}}(B_s\to\mu^+\mu^-)$ turn out to be small.
\item
On the other hand if $B_d\to K^*\mu^+\mu^-$ anomaly disappeared but 
future more precise data would definitely show that  $\overline{\mathcal{B}}(B_s\to\mu^+\mu^-)$ is significantly below its SM value, other models, in particular the one with $\beta=-2/\sqrt{3}$ but fermion representations $F_2$, would be 
favoured. Further tests would come from future measurements of decays 
with neutrinos in the final state.
\item
As seen in Figs.~\ref{fig:BsmuRnu} and  \ref{fig:BsmuRnuF2} in the case 
of neutrinos in the final state the dependence on $\tan\bar\beta$ is opposite 
to the case of $B_{s,d}\to\mu^+\mu^-$ and $Z$ and $Z^\prime$ contributions 
interfere constructively for small $\tan\bar\beta$ but for large $\tan\bar\beta$ 
they cancel each other to a large extent. 
\item
There is no specific correlations between the branching ratios for 
$B_s\to\mu^+\mu^-$ and $B_d\to\mu^+\mu^-$ decays and this implies
significant departures from CMFV relation between their branching ratios.
We show as an example in Fig.~\ref{fig:BdvsBs} the results for values of $\tan\bar\beta=1,~ 5,~0.2$ and 
 fermion representations $F_1$. The case without $Z-Z^\prime$ mixing, that is 
pure $Z^\prime$ contributions, represented by the black regions allows to 
see that the presence of 
$Z$ boson contributions in both decays represented by departure from 
these areas can be  significant as could be deduced from previous results.
\item
As seen in Fig.~\ref{fig:c10} $Z$ boson contributions have significant 
effect on the size of CP violation in $B_s\to\mu^+\mu^-$  that originates from a non-vanishing imaginary part of $C_{10}^{NP}$, in constrast to $C^{SM}_{10}$ that is real.  As the tests 
of CP-violating effects in this decay are in the distant future we only show 
the results for $F_1$ in this case.
\end{itemize}

{ Finally, we look at the $B_d\to K\mu^+\mu^-$ decay and its correlation with 
 $\RE(C_9^\text{NP})$. The interest in the analysis of this decay lies 
in the fact that in contrast to $B_s\to\mu^+\mu^-$ and  $B_d\to K^*\mu^+\mu^-$ 
that in 331 models are sensitive only to  $C_{10}^\text{NP}$ and  $C_9^\text{NP}$, respectively, 
the branching ratio for  $B_d\to K\mu^+\mu^-$ depends on both coefficients.
 Moreover, lattice calculations of the relevant form factors 
are making significant progress here \cite{Bouchard:2013mia,Bouchard:2013eph} 
and the importance of this decay will increase in the future.

Neglecting the interference between NP contributions
the formula for the differential branching ratio confined to 
large $q^2$ region ($15\gev^2\le q^2\le 22\gev^2$)\footnote{This formula is based on 
\cite{Altmannshofer:2013foa} and a recent update. Straub, private communication.}
 reduces in 331 models in units  of $1/\gev^2$  to 
\be\label{LHSK}
10^9\times \frac{d\mathcal{B}(B_d\to K\mu^+\mu^-)_{[15,22]}}{d q^2}=13.1+ 
3.15~\RE(C_9^\text{NP})-3.23\RE(C_{10}^\text{NP}),
\ee
The relevant Wilson coefficients are given in (\ref{C9}) and (\ref{C10}). 
This formula describes triple correlation 
 between  $B_d\to K\mu^+\mu^-$,  
$\RE(C_9^{\rm NP})$  and $\overline{\mathcal{B}}(B_s\to\mu^+\mu^-)$ which constitutes an important test for the models in question.  
In the absence of $Z-Z^\prime$ mixing this triple correlation involving the
rate for $B_d\to K\mu^+\mu^-$ at large $q^2$ can be found  
in Fig.~13 in \cite{Buras:2013dea} for the case of $F_1$ representations. 

The error on the first SM term is estimated to be $10\%$ \cite{Bouchard:2013mia,Bouchard:2013eph}. This should 
 be compared with the LHCb result \cite{PatelMoriond}
\be\label{equ:expBdKmu}
10^9\times \frac{d\mathcal{B}(B_d\to K\mu^+\mu^-)_{[15,22]}}{d q^2}=12.1\pm 0.4\pm 0.6 \qquad {(\rm LHCb).}
\ee

\begin{figure}[!tb]
 \centering
\includegraphics[width = 0.44\textwidth]{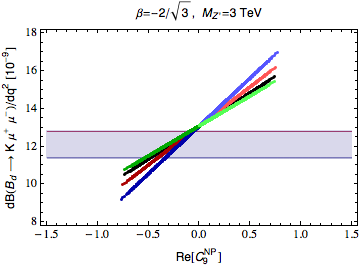}
\includegraphics[width = 0.44\textwidth]{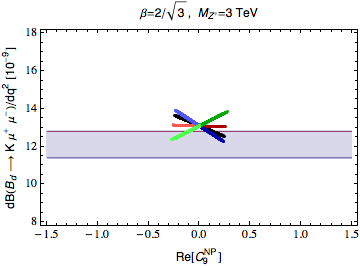}
\\
\includegraphics[width = 0.44\textwidth]{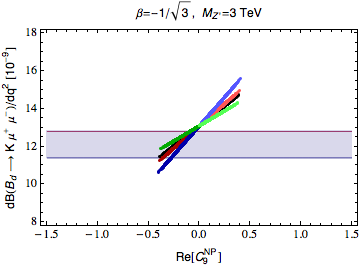}
\includegraphics[width = 0.44\textwidth]{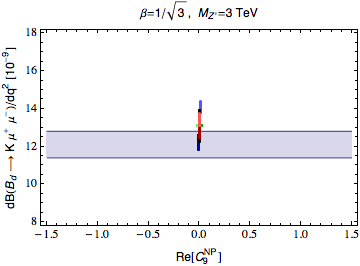}
\caption{Correlation of the differential branching ratio for 
 $B_d\to K\mu^+\mu^-)_{[15,22]}$  versus $\RE(C_9^{\rm NP})$ for
$\beta=\pm 1/\sqrt{3}$ and $\beta=\pm 2/\sqrt{3}$ and $F_1$ setting $M_{Z^\prime}=3\tev$ 
and different values of $\tan\bar\beta$ with colour coding given in 
(\ref{tanbetacoding}).
$\overline{\mathcal{B}}(B_s\to\mu^+\mu^-)\leq\overline{\mathcal{B}}(B_s\to\mu^+\mu^-)_\text{SM}$ (darker colours) and 
$\overline{\mathcal{B}}(B_s\to\mu^+\mu^-)\geq\overline{\mathcal{B}}(B_s\to\mu^+\mu^-)_\text{SM}$ (lighter colours) . The gray regions show  the 
experimental range in~(\ref{equ:expBdKmu}).
}\label{fig:pBsmuvsBdKmu}~\\[-2mm]\hrule
\end{figure}

\begin{figure}[!tb]
 \centering
\includegraphics[width = 0.44\textwidth]{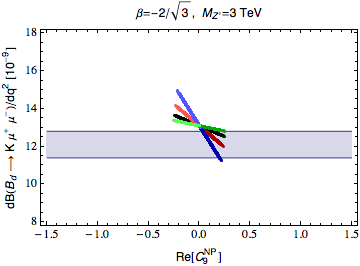}
\includegraphics[width = 0.44\textwidth]{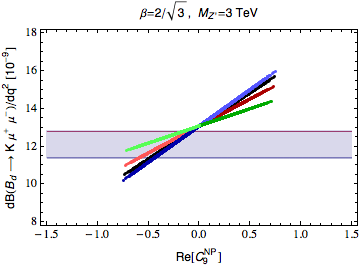}
\\
\includegraphics[width = 0.44\textwidth]{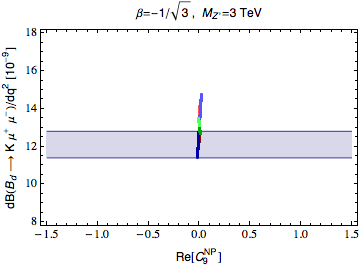}
\includegraphics[width = 0.44\textwidth]{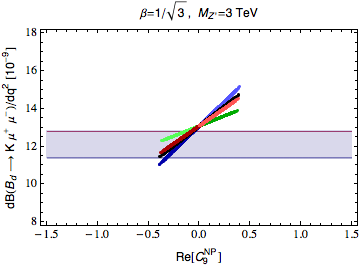}
\caption{As in Fig.~\ref{fig:pBsmuvsBdKmu} but for $F_2$.
}\label{fig:pBsmuvsBdKmuF2}~\\[-2mm]\hrule
\end{figure}

In Fig.~\ref{fig:pBsmuvsBdKmu} we generalize this result to include 
$Z-Z^\prime$ mixing and in Fig.~\ref{fig:pBsmuvsBdKmuF2} we present corresponding results for 
$F_2$. As in the case of Figs.~\ref{fig:ReC9vsBsmu} and \ref{fig:ReC9vsBsmuF2} 
we obtain straight lines with slopes depending on the values of $\beta$ and 
$\tan\bar\beta$. The case of no $Z-Z^\prime$ mixing is again shown by black 
lines. In these plots the colour coding is:
\be\label{tanbetacoding}
\tan\bar\beta= 1.0~\text{(red)},\quad \tan\bar\beta= 5.0~\text{(blue)},\quad 
\tan\bar\beta= 0.2~\text{(green)},
\ee
with lighter colours showing when  $\overline{\mathcal{B}}(B_s\to\mu^+\mu^-)$ 
is enhanced and with darker ones when it is suppressed with respect to the 
SM prediction.

There are two striking differences between these results an those in
Figs.~\ref{fig:ReC9vsBsmu}-- \ref{fig:BsmuRnuF2}. The effects of $Z-Z^\prime$ 
mixing  in Figs.~\ref{fig:pBsmuvsBdKmu} and \ref{fig:pBsmuvsBdKmuF2} are 
significantly smaller and consequently the symmetry in (\ref{symmetry}) is 
less broken.}

 At this point we would like to emphasize that all the results  described until 
now  do not take into account the constraints from electroweak precision tests. 
In Section~\ref{EWP} we will analyze which lines in Figs.~\ref{fig:ReC9vsBsmu}-- \ref{fig:BsmuRnuF2}, \ref{fig:pBsmuvsBdKmu} and \ref{fig:pBsmuvsBdKmuF2} survive the latter tests and which not. In any case 
with $M_{Z^\prime}=3\tev$, as demonstrated in  \cite{Buras:2013dea}, the bounds from LEP-II and LHC are satisied.

\boldmath
\subsection{Numerical Results for Rare $K$ Decays}
\unboldmath
In our recent paper \cite{Buras:2014sba} we have reemphasized the strong dependence of rare $K$ decay branching ratios on the values of the elements  of the 
CKM matrix $\vub$ and $\vcb$. This dependence is particularly strong 
in the case of $\klpn$ as seen in Table~3 of that paper. While in \cite{Buras:2014sba} we have studied six scenarios for $\vub$ and $\vcb$ in 331 models 
most of these scenarios are ruled out by $\varepsilon_K$. In fact as 
already pointed out in \cite{Buras:2012dp} NP effects in  $\varepsilon_K$ 
are rather small when constraints from $B_{d,s}^0-\bar B_{d,s}$ mixing are taken 
into account.  Therefore the 331 models can only be made consistent 
with data on $\varepsilon_K$ for values of 
 $\vub$ and $\vcb$ for which the SM prediction for $\varepsilon_K$ is rather 
close to this data. Then only scenarios $d)$ and $f)$ in  \cite{Buras:2014sba}
\begin{align}
 d)&\qquad \vub = 4.1\times 10^{-3}\qquad \vcb = 42.0\times 10^{-3} \label{sd}\\
f)& \qquad \vub = 3.9\times 10^{-3}\qquad \vcb = 42.0\times 10^{-3}  \label{sf}
\end{align}
survive the $\varepsilon_K$ constraint in 331 models as then for central values  of remaining parameters $|\varepsilon_K|=2.35\times 10^{-3}$ and 
$|\varepsilon_K|=2.25\times 10^{-3}$, respectively. This is close to 
the experimental value $|\varepsilon_K|=2.23\times 10^{-3}$ so that there 
is no problem in fitting the data.

\begin{figure}[!tb]
 \centering
\includegraphics[width = 0.45\textwidth]{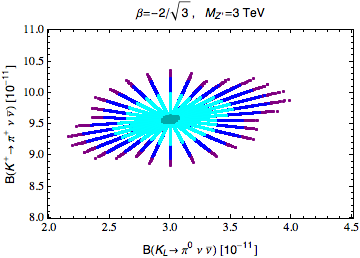}
\includegraphics[width = 0.45\textwidth]{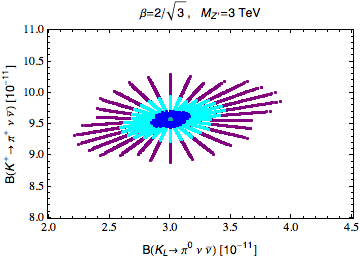}

\includegraphics[width = 0.45\textwidth]{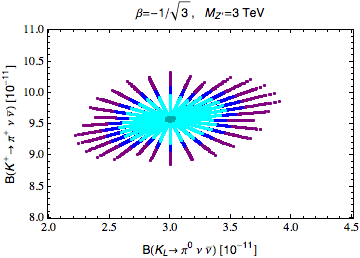}
\includegraphics[width = 0.45\textwidth]{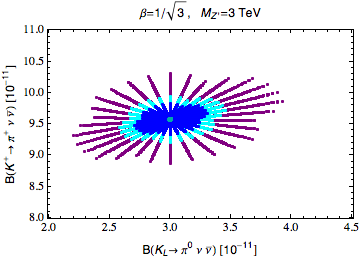}

\caption{\it $\kpn$ versus $\klpn$ for different values of $\beta$ 
for $\tan\bar\beta=1.0$ (cyan), $\tan\bar\beta=5.0$ (darker cyan), $\tan\bar\beta = 0.2$ (purple)  and without $Z-Z^\prime$ mixing (blue) for 
$F_1$ representations. 
}\label{fig:KCKMad}~\\[-2mm]\hrule
\end{figure}

\begin{figure}[!tb]
 \centering
\includegraphics[width = 0.45\textwidth]{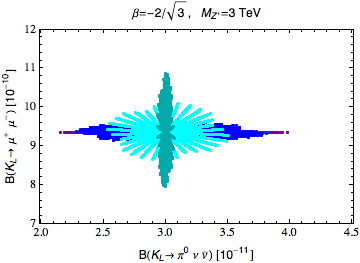}
\includegraphics[width = 0.45\textwidth]{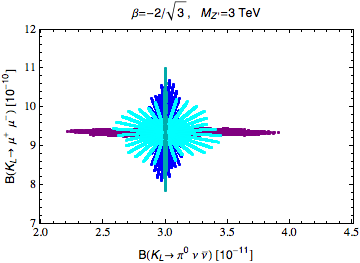}

\includegraphics[width = 0.45\textwidth]{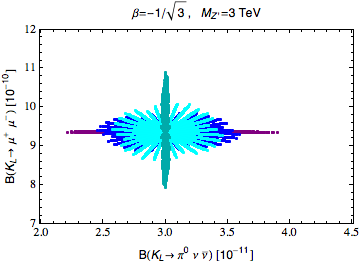}
\includegraphics[width = 0.45\textwidth]{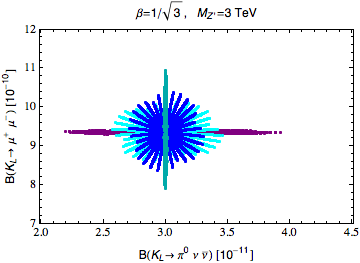}
\caption{\it $\mathcal{B}(K_L\to\mu^+\mu^-)$ versus $\klpn$ for different values of $\beta$ for $\tan\bar\beta=1.0$ (cyan), $\tan\bar\beta=5.0$ 
(darker cyan), $\tan\bar\beta = 0.2$ (purple) and without $Z-Z^\prime$ mixing (blue) for $F_1$ representations. 
}\label{fig:KLCKMad}~\\[-2mm]\hrule
\end{figure}

\begin{figure}[!tb]
 \centering
\includegraphics[width = 0.45\textwidth]{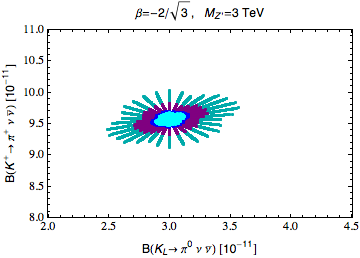}
\includegraphics[width = 0.45\textwidth]{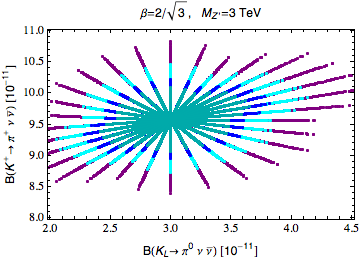}

\includegraphics[width = 0.45\textwidth]{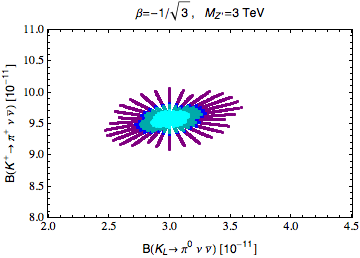}
\includegraphics[width = 0.45\textwidth]{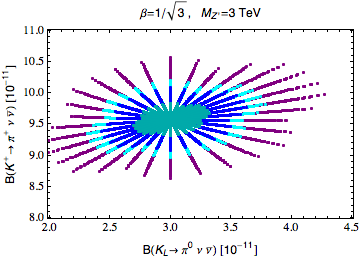}

\caption{\it $\kpn$ versus $\klpn$ as in Fig.~\ref{fig:KLCKMad} but for F2.
}\label{fig:KCKMadF2}~\\[-2mm]\hrule
\end{figure}

Now in \cite{Buras:2013dea} and in the analysis of $B$ decays in the present paper we have used the values
\be
\vub = 3.6\times 10^{-3}\qquad \vcb = 42.4\times 10^{-3} 
\ee
which implies $|\varepsilon_K|=2.17\times 10^{-3}$ that is rather close to 
the choice $f)$ above and as it also allows to satisfy the $\varepsilon_K$ 
constraint, we will use these values for rare $K$ decays and $\epe$.

In Figs.~\ref{fig:KCKMad} and \ref{fig:KLCKMad} we show various correlations 
between rare decay branching ratios for the four 331 models considered for the 
$F_1$ case, three values of $\tan\bar\beta=1,5,0.2$ and the case without $Z-Z^\prime$ mixing which represents sole $Z^\prime$ contributions. While these plots are 
self-explanatory, in particular when considered simultaneously with 
Table~\ref{tab:Rmumu}, we would like to emphasize a number of most important 
features in them. These are:
\begin{itemize}
\item
A rather striking feature is the cancellation of $Z^\prime$ and $Z$ contributions to the branching ratios for $\kpn$ and $\klpn$ for $\tan\bar\beta=5.0$ so 
that in this case one obtains basically the SM prediction 
independently of the value of $\beta$. On the contrary  in the case of $K_L\to\mu^+\mu^-$  for $\tan\bar\beta=5.0$ and in particular for negative $\beta$ NP effects are enhanced through 
$Z-Z^\prime$ mixing. This difference between decays with neutrino and muons 
in the final state has been also seen in the case of $B$ decays.
\item
A different behaviour is observed for $\tan\bar\beta=1.0$. 
In the case of $\kpn$ and $\klpn$ for models 
with $\beta=-2/\sqrt{3}$ and  $\beta=-1/\sqrt{3}$ there is a {\it destructive} 
interference between $Z$ and $Z^\prime$ contributions decreasing somewhat 
NP effects due to $Z^\prime$ exchange alone. For  $\beta=2/\sqrt{3}$ and  $\beta=1/\sqrt{3}$, on the other hand, the corresponding interference is {\it constructive} and NP effects are increased. On the contrary the effects of $Z-Z^\prime$ 
mixing for $\tan\bar\beta=1.0$ in $K_L\to\mu^+\mu^-$ are rather small.
\item
For $\tan\bar\beta=0.2$ NP effects in $\klpn$ and $\kpn$ increase but they 
basically disappear in the case $K_L\to\mu^+\mu^-$.
\item
On the whole NP effects except for the case of $\klpn$  and 
$\tan\bar\beta\le 1$ are rather small and 
it will be difficult to distinguish them from SM expectations unless 
parametric uncertainties decrease by much and experimental data will be very 
precise.
\end{itemize}

 { Considering the case of $F_2$ representations one can deduce from Table~\ref{tab:Rmumu2} that the symmetry in (\ref{symmetry}) is significantly broken 
in rare $K$ decays but the size of NP effects is similar to the $F_1$ case.
As an example we show in Fig.~\ref{fig:KCKMadF2}  the correlation between the branching ratios for $\kpn$ and $\klpn$ for $F_2$. In particular for $\beta=2/\sqrt{3}$ and $\tan\bar\beta=0.2$ NP effects are significant.}

But the main message from our analysis of rare $B$ and $K$ decays 
is that neglecting $Z$ contributions 
in decays governed by axial vector couplings to muons or left-handed 
couplings to neutrinos is not justified and observing significant NP
effects in $B_{s,d}\to\mu^+\mu^-$ would imply $\tan\bar\beta>1$ and only small 
effects in $\klpn$ and $\kpn$. On the other hand confirming SM predictions for 
$B_{s,d}\to\mu^+\mu^-$ to high degree would in 331 models for $M_{Z^\prime}$ 
still allow for modest by significant departures from SM expectations for 
these decays and imply for 331 models $\tan\bar\beta\le 1$.

 Again the results just described 
 do not take into account the constraints from electroweak precision tests 
and it will be interesting to see in  Section~\ref{EWP} the impact of the latter tests on them.

\boldmath
\section{The Ratio $\epe$}\label{sec:4}
\unboldmath
\subsection{Preliminaries}
Recently we have presented a new analysis of $\epe$ within the SM and models 
with tree-level $Z^\prime$ and $Z$ boson exchanges \cite{Buras:2014sba}. Several of the formulae 
presented in that paper can be directly used in the context of the 331 models 
and consequently our presentation will be rather brief. In 331 models we 
have
\be\label{total}
\left(\frac{\varepsilon'}{\varepsilon}\right)_{331}=\left(\frac{\varepsilon'}{\varepsilon}\right)_{\rm SM}+\left(\frac{\varepsilon'}{\varepsilon}\right)_{Z}+
\left(\frac{\varepsilon'}{\varepsilon}\right)_{Z^\prime}\, ,
\ee
where the formula for the SM contribution, an update of the original one in 
\cite{Buras:2003zz}, is given in (53) of \cite{Buras:2014sba}.
\boldmath
\subsection{$Z$ Contribution}
\unboldmath
This case is simple as the only thing to be done is to introduce shifts
in the functions $X$, $Y$ and $Z$ that enter the SM model contribution to $\epe$. 
Using the results of Section 7 in 
\cite{Buras:2014sba} together with the relation (\ref{ZZprime}) we 
find
\be
\Delta X=\Delta Y =\Delta Z= \sin\xi \, c_W\frac{8\pi^2}{g^3}\frac{{\rm Im}\Delta_L^{sd}(Z^\prime)}{{\rm Im}\lambda_t}
\ee
where $g=0.652$ and $\lambda_t=V_{td}V^*_{ts}$. Replacing then the functions 
$X_0(x_t)$, $Y_0(x_t)$ and $Z_0(x_t)$ by 
\be
X=X_0(x_t)+\Delta X, \qquad Y=Y_0(x_t)+\Delta Y, \qquad Z=Z_0(x_t)+\Delta Z
\ee
in the 
formula (53) for $\epe$ in \cite{Buras:2014sba} allows to take automatically the first two 
contributions in (\ref{total})  in 331 models into account.

\boldmath
\subsection{$Z^\prime$ Contribution}
\unboldmath
Using the general formulae for the flavour diagonal $Z^\prime$ couplings to 
quarks in \cite{Buras:2012dp,Buras:2013dea}
we find that in LO as far as penguin operators are concerned 
the only non-vanishing Wilson coefficients at 
$\mu=M_{Z^\prime}$ are the ones of the known QCD penguin operator $Q_3$, the known
electroweak penguin operator $Q_7$ as well as of the operator 
\be
\tilde Q_3= (\bar s d)_{V-A}\left[(\bar bb)_{V-A} + (\bar tt)_{V-A}\right]
\ee
which is present due to different couplings of $Z^\prime$ to the third generation of quarks.  Their coefficients are given in 331 models as follows
\be
C_3(M_{Z^\prime})= \frac{g}{2\sqrt{3}c_W}\sqrt{f(\beta)} \left[-1+(1+\frac{\beta}{\sqrt{3}})s_W^2\right] \frac{\Delta_L^{sd}(Z^\prime)}{4 M_{Z^\prime}^2}
\ee
\be
C_7(M_{Z^\prime})= \frac{g}{2\sqrt{3}c_W}\sqrt{f(\beta)}\frac{4}{\sqrt{3}}\beta s_W^2\frac{\Delta_L^{sd}(Z^\prime)}{4 M_{Z^\prime}^2}
\ee
\be
\tilde C_3(M_{Z^\prime})= \frac{g}{2\sqrt{3}c_W}\sqrt{f(\beta)} \left[2 c_W^2\right] \frac{\Delta_L^{sd}(Z^\prime)}{4 M_{Z^\prime}^2}
\ee
with $f(\beta)$ defined in (\ref{central}). We recall that these results are 
valid for fermion representation $F_1$. For $F_2$ one just has to reverse the 
sign in front of $\beta$ according to the rules outlined in Section~\ref{Frepresentation}.

The coefficients of these three operators at $\mu=M_{Z^\prime}$  are of the same order. Yet, when QCD renormalization group effects are included and the size of 
hadronic matrix elements relevant for $\epe$ are taken into account we find
that the $Q_3$ and $\tilde Q_3$ contributions can be neglected leaving 
the left-right electroweak penguin operators 
\begin{equation}\label{O4} 
Q_7 = \frac{3}{2}\;(\bar s d)_{V-A}\sum_{q=u,d,s,c,b,t}e_q\;(\bar qq)_{V+A} 
~~~~~ Q_8 = \frac{3}{2}\;(\bar s_{\alpha} d_{\beta})_{V-A}\sum_{q=u,d,s,c,b,t}e_q
        (\bar q_{\beta} q_{\alpha})_{V+A}\, 
\end{equation}
as the only relevant operators 
in $Z^\prime$ contribution 
to $\epe$ in 331 models. 
However, even if at LO the Wilson 
coefficient of $Q_8$ vanishes at $\mu=M_{Z^\prime}$,  at $\mu=m_c$ used 
in our numerical analysis of $\epe$ its Wilson coefficient $C_8(m_c)$ is of the 
same order as $C_7(m_c)$. Indeed 
the relevant 
one-loop anomalous dimension matrix in the $(Q_7,Q_8)$ basis is given by  
\begin{equation}
\hat \gamma^{(0)}_s = 
\left(
\begin{array}{cc}
2 & -6  \\ \svs
0 & -16
   \end{array}
\right).
\label{reducedZ}
\end{equation}

Performing the renormalization group evolution from $M_{Z^\prime}$ to $m_c=1.3\gev$ we find using explicit formulae in \cite{Buras:2014sba}
\be\label{LOC7C8}
 C_7(m_c)= 0.82\, C_7(M_{Z^\prime})\qquad   C_8(m_c)= 1.35\,  C_7(M_{Z^\prime}).
\ee

Due to the large element $(1,2)$ in the matrix (\ref{reducedZ}) and 
the large anomalous dimension of the $Q_8$ operator represented by the $(2,2)$ 
element in (\ref{reducedZ}), the two coefficients are comparable in size.
But  the matrix element $\langle Q_7\rangle_{2}$ is colour suppressed 
which is not the case of  $\langle Q_8\rangle_{2}$ and within a good 
approximation we can neglect the contributions of $Q_7$. In summary, it 
is sufficient to keep only $Q_8$ contributions in $\epe$ in 331 models.

The relevant isospin amplitude $A_2$ in $K\to\pi\pi$ decays necessary to 
calculate $Z^\prime$ contribution to $\epe$  is 
given as follows
\be\label{RENPZ2}
A_2^{\rm NP}= C_8(m_c)\langle Q_8 (m_c)\rangle_2
\ee
where \cite{Buras:2014sba}
\be
\langle Q_8(m_c)\rangle_2 =0.57 \, \left[ \frac{114\mev}{m_s(m_c) + m_d(m_c)}\right]^2 \,\left[\frac{B_8^{(3/2)}}{0.65}\right]\,\gev^3\, 
\ee
with
$\bei= 0.65\pm 0.05$ from lattice QCD \cite{Blum:2012uk}.

In our numerical analysis 
we will use for the quark masses the values from FLAG 2013  
\cite{Aoki:2013ldr}
\be
m_s(2\gev)=(93.8\pm2.4) \mev, \qquad
m_d(2\gev)=(4.68\pm0.16)\mev.
\ee
Then at the nominal value $\mu=m_c=1.3\gev$ we have
\be
m_s(m_c)=(108.6\pm2.8) \mev, \qquad
m_d(m_c)=(5.42\pm 0.18)\mev.
\ee

The final expression for $Z^\prime$ contributions is given by
\be\label{eprimeZfinal}
\left(\frac{\varepsilon'}{\varepsilon}\right)_{Z^\prime}=
\frac{\omega_+}{|\varepsilon_K|\sqrt{2}}\frac{{\IM} A_2^{\rm NP}}{{\RE}A_2}
\ee
where  \cite{Cirigliano:2003nn}
\be
\omega_+=(4.1\pm0.1)\times 10^{-2}\,.
\ee
In evaluating (\ref{eprimeZfinal}) we use, as in the case of the SM, the experimental values for  ${\rm Re} A_2$ and $\varepsilon_K$:
\be
{\rm Re}A_2= 1.210(2)   \times 10^{-8}~\gev,\qquad
|\varepsilon_K|=2.228(11)\times 10^{-3}\,\,.
\ee

As NP contributions to $\epe$, both due to $Z$ and $Z^\prime$, are dominated 
by the operator $Q_8$, the ratio of these contributions depends with high accuracy simply on the ratio of $C_8(m_c)$ in these two contributions. This allows 
then to derive a simple relation
\be\label{equ:Repsdef}
\left(\frac{\varepsilon'}{\varepsilon}\right)_{Z}=R_{\varepsilon^\prime}
\left(\frac{\varepsilon'}{\varepsilon}\right)_{Z^\prime},
\ee
where ($s_W^2=0.23116$)
\be\label{equ:Repsprime}
R_{\varepsilon^\prime}=-\frac{0.53}{\beta}\, \frac{c_W^2}{3}\left[3\beta\frac{s_W^2}{c_W^2}+\sqrt{3}{a}\right] =-\frac{0.14}{\beta}\, \left[0.90\beta+\sqrt{3}a\right].
\ee
This expression is valid for fermion representation $F_1$. For $F_2$ for a given $\beta$ one should just remove the overall minus sign in this expression as the $\beta$ in front of the parenthesis enters the $Z^\prime$ coupling. The 
expression in the parenthesis comes from $Z-Z^\prime$ mixing and is independent 
of the fermion  representation.
The factor $0.53$ summarizes the difference in renormalization group effects 
for $Z$ and $Z^\prime$ contributions. QCD renormalization of $Q_8$ gives alone 
$0.56$ and additional small suppression comes from the running of electroweak 
parameters. We list the values of $R_{\varepsilon^\prime}$ in the last row of 
 Tables~\ref{tab:Rmumu}-\ref{tab:Rmumu3}. Evidently $Z^\prime$ dominates 
NP contributions to $\epe$ implying that $Z-Z^\prime$ mixing effects are 
 small in this ratio. The two exceptions are the case of $\beta=-1/\sqrt{3}$ 
and $\tan\bar\beta=5$ and the  case of $\beta=1/\sqrt{3}$ 
and $\tan\bar\beta=0.2$ for which $Z$ contribution reaches $50\%$ of the $Z^\prime$ 
one.

\boldmath
\subsection{Numerical Analysis of $\epe$}
\unboldmath
\begin{figure}[!tb]
 \centering
\includegraphics[width = 0.45\textwidth]{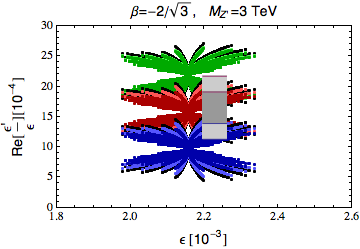}
\includegraphics[width = 0.45\textwidth]{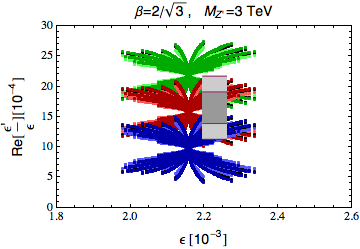}

\includegraphics[width = 0.45\textwidth]{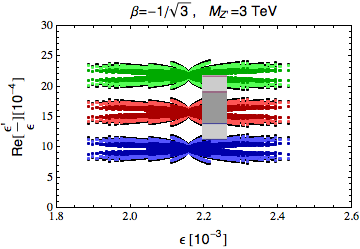}
\includegraphics[width = 0.45\textwidth]{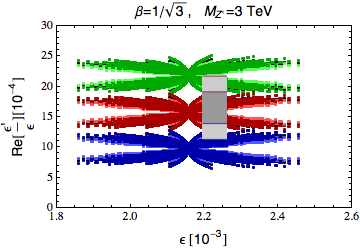}
\caption{\it $\epe$ versus $\varepsilon_K$, $M_{Z^\prime} = 3~$TeV, $F_1$ representations, different values of $\beta = 
\pm\frac{2}{\sqrt{3}},\pm\frac{1}{\sqrt{3}}$, $\tan\bar\beta = 1(5)$ lighter colours (darker colours) and no $Z-Z^\prime$ mixing (black) 
and $B_6^{(1/2)} = 0.75$ (blue), 
$B_6^{(1/2)} = 
1.00$ (red), $B_6^{(1/2)} = 1.25$ (green). (Light) gray area: $1\sigma$($2\sigma$) range of $\epe$ and $3\sigma$ range of $\varepsilon_K$.
}\label{fig:epevsepsKCKMp}~\\[-2mm]\hrule
\end{figure}

\begin{figure}[!tb]
 \centering
\includegraphics[width = 0.45\textwidth]{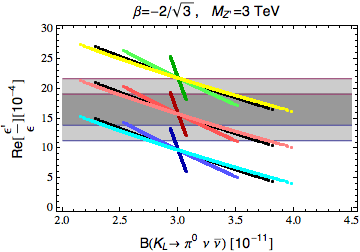}
\includegraphics[width = 0.45\textwidth]{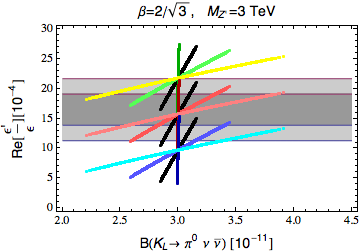}

\includegraphics[width = 0.45\textwidth]{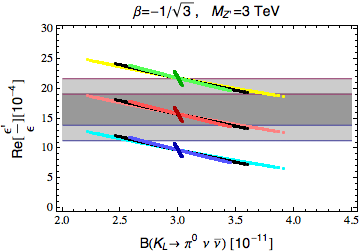}
\includegraphics[width = 0.45\textwidth]{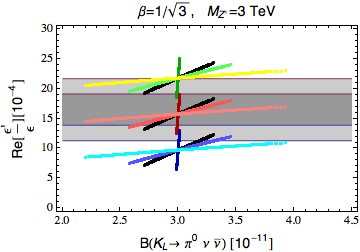}
\caption{\it $\epe$ versus $\mathcal{B}(\klpn)$, $M_{Z^\prime} = 3~$TeV, different values of $\beta = 
\pm\frac{2}{\sqrt{3}},\pm\frac{1}{\sqrt{3}}$, different values of $B_6^{(1/2)} = 0.75,~1.00,~1.25$ and  $\tan\bar\beta = 1$ (light blue, light red, light green), $\tan\bar\beta = 5$ 
(dark blue, dark red, dark green), $\tan\bar\beta = 0.2$ 
(cyan, pink, yellow) and no $Z-Z^\prime$ mixing (black, black, black). Fermion representations $F_1$.
}\label{fig:Br0vsepsKCKMp}~\\[-2mm]\hrule
\end{figure}

In Fig.~\ref{fig:epevsepsKCKMp} we show $\epe$ versus $\varepsilon_K$. 
We make the following observations:
\begin{itemize}
\item
331 models for all values of $\beta$ are consistent with the data for $\epe$ 
provided 
 $\bsi\approx 1.0$ represented by {\it red} colour.
\item
$Z-Z^\prime$ mixing effects are for most parameters 
small as already expected on the basis of 
the formula (\ref{equ:Repsprime}) and the values of $R_{\varepsilon^\prime}$ in 
Tables~\ref{tab:Rmumu}-\ref{tab:Rmumu3}.
\end{itemize}

In Fig.~\ref{fig:Br0vsepsKCKMp} we show  $\epe$ versus $\mathcal{B}(\klpn)$ for  fermion representations $F_1$. We observe that the correlation between these two observables is very strict 
as $\epe$ is linear in the imaginary parts of the flavour violating $Z^\prime$ and $Z$ couplings and NP contribution to $\mathcal{B}(\klpn)$ is dominated by 
the interference of the SM and NP amplitude. Consequently is also linear in 
these 
couplings. Two interesting features of these plots should be emphasized
\begin{itemize}
\item
 while for {\it negative} $\beta$ the ratio
$\epe$ decreases with increasing $\mathcal{B}(\klpn)$, 
 for {\it positive} $\beta$ it increases with increasing $\mathcal{B}(\klpn)$. 
The latter property is rather rarly  found in other extensions of the SM. 
\item
The effects of $Z-Z^\prime$ mixing are clearly visible, in particular for 
$\beta>0$. They originate dominantly in the ones present in $\klpn$.
\item
As already known from Fig.~\ref{fig:epevsepsKCKMp} the agreement of the theory 
and the data is best for $\bsi\approx 1.0$. 
\end{itemize}

\section{Electroweak Precision Observables}\label{EWP}
\subsection{Preliminaries}
The modifications of the $Z$ boson couplings due to $Z-Z^\prime$ mixing in 331 
models can be tested through electroweak precision measurements at LEP-I and 
SLD and such an analysis can be found in  \cite{Ochoa:2005ih} and recently in \cite{Richard:2013xfa}. As our formula for $\sin\xi$ 
differs from the one used in these papers we want to analyze the 
impact of $Z-Z^\prime$ mixing on most important electroweak precision observables (EWPO) and study their dependence on $\beta$, $\tan\bar\beta$ and the choice of 
fermion representations. Thus our next goal is to construct three tables for 
the shifts of a number of EWPO relative to SM predictions. This will eventually allow 
us to investigate the correlations between NP effects in EWPO and flavour 
observables. In these tables we will set $M_{Z^\prime}=3\tev$. As the shifts 
in question are inversely proportional to  $M_{Z^\prime}^2$ it is straight 
forward to find out what happens for other values of $M_{Z^\prime}$.

Transparent analyses  of the effects of $Z-Z^\prime$ mixing in EWPO can 
be found in \cite{Altarelli:1990dt,Altarelli:1996pr,Chiappetta:1996km}. 
We will follow here the general analysis in \cite{Altarelli:1996pr} and to this 
end it is useful to note that our couplings of a given fermion $f$ to $Z$ and $Z^\prime$  differ from the couplings $v_{S,N}^f$ and $a_{S,N}^f$ used in 
that paper by an overall factor:
\be
\Delta_V^{ff}(Z)=\frac{g}{\cos\theta_W} v_S^f, \qquad 
\Delta_A^{ff}(Z)=\frac{g}{\cos\theta_W} a_S^f,
\ee
\be
\Delta_V^{ff}(Z^\prime)=\frac{g}{\cos\theta_W} v_N^f, \qquad 
\Delta_A^{ff}(Z^\prime)=\frac{g}{\cos\theta_W} a_N^f\,.
\ee
For $Z$ couplings we then have
\be
v_S^f=T^f_{3L}-2\sin^2\theta_W Q^f, \qquad a_S^f=-T^f_{3L}
\ee
where $T^f_{3L}$ is the third component of the weak isospin of the fermion $f$ 
and $Q^f$ its electric charge. Our effective $Z$ couplings in (\ref{effZ})
are then related to the analogously defined  ones in  \cite{Altarelli:1996pr} simply as 
follows 
\be
[\Delta^{f}_V(Z)]_{\rm eff}= \frac{g}{\cos\theta_W}\,v^f_{\rm eff}, \qquad [\Delta^{f}_A(Z)]_{\rm eff}=\frac{g}{\cos\theta_W}\, a^f_{\rm eff}.
\ee
 
Due to the $Z-Z^\prime$ mixing the $\varrho$ parameter, defined by 
\be\label{varrho}
\varrho=\frac{M_W^2}{M_Z^2\cos^2\theta_W}
\ee
receives tree-level contribution $\Delta\varrho_M$ which for $M_{Z^\prime}\gg M_Z$ 
is given by \cite{Altarelli:1990dt,Altarelli:1996pr,Chiappetta:1996km}
\be
\Delta\varrho_M=\left(\frac{M^2_{Z^\prime}}{M^2_Z}\right)\sin^2\xi\, .
\ee
This shift is strictly positive and $\ord(M^2_Z/M^2_{Z^\prime})$. Consequently it is  of the 
same order as the $Z-Z^\prime$ mixing effects in  the effective $Z$ couplings in (\ref{effZ}) so that it has to be taken into account. 
 On the other hand we conclude on the basis of \cite{Hoang:1999yv} that in 331 models studied by us the 
oblique contributions involving new heavy charged gauge bosons, scalars and 
fermions can be neglected when their masses are in the few TeV regime. Consequently the shift in $\Delta\varrho$ due to NP in these models 
is dominated by $\Delta\varrho_M$ given above.

Keeping fixed as input parameters $\alpha(M_Z)$, $G_F$ and $M_Z$, the effective 
Weinberg angle in~(\ref{varrho}) is modified due to the shift in $\varrho$ 
as follows \cite{Altarelli:1990dt,Altarelli:1996pr}
\be\label{swshift}
\Delta(\sin^2\theta_W)=-\kappa\Delta\varrho_M, \qquad \kappa=\frac{\sin^2\theta_W \cos^2\theta_W}{\cos^22\theta_W}\,.
\ee

It should be noted that this shift is strictly negative and this property 
is valid for any $Z^\prime$ model. Yet, as we will see below, this does 
not mean that the shift in the so-called effective $\sin^2\theta^l_\text{eff}$
 is negative in any $Z^\prime$ model as the contributions to 
$Z$ couplings that are proportional to $\sin\xi$ are clearly model 
dependent. Some aspects of this dependence has been already analyzed 
in the context of 331 models in \cite{Richard:2013xfa}. 
Our improved formula 
for $\sin\xi$ in (\ref{sxi}) and the gained insight on the correlations 
between $Z-Z^\prime$ mixing and FCNCs processes allows us to have another 
look at this issue thereby also generalizing the analysis in \cite{Richard:2013xfa}.

Next denoting a given observable by $\mathcal{O}$ the shift due to $Z-Z^\prime$  mixing can be linearized in $\Delta\varrho_M$ and $\sin\xi$ as follows
 \cite{Altarelli:1996pr,Chiappetta:1996km}
\be\label{ALT}
\frac{\delta\mathcal{O}}{\mathcal{O}}=A_{\mathcal{O}}\Delta\varrho_M+
B_{\mathcal{O}}\sin\xi\,.
\ee
Here the coefficients $A_{\mathcal{O}}$ are universal and depend only 
on the SM parameters and couplings. On the other hand $B_{\mathcal{O}}$ 
depend on the diagonal $Z^\prime$-couplings in (\ref{effZ}). { Direct 
$Z^\prime$ contributions to EWPO are negligible.}

We will list the general formulae for the coefficients  $A_{\mathcal{O}}$ and $B_{\mathcal{O}}$  below but rather then giving numerical values for them 
as done in \cite{Altarelli:1996pr} we will calculate the shifts in 
EWPO for different values of $\beta$, $\tan\bar\beta$ and the fermion 
representations $F_1$ and $F_2$ for $M_{Z^\prime}=3\tev$. To this end 
we will proceed as follows
\begin{itemize}
\item
We will calculate a number of EWPO 
as functions of the effective diagonal $Z$-couplings in (\ref{effZ})
in 331 models using tree-level formulae. In doing this one should remember 
that in addition to the direct contribution of $Z^\prime$ to the effective $Z$ 
couplings leading to the second term in (\ref{ALT}) also the shift in $\sin^2\theta_W$ entering the vector $Z$ coupling has to be taken into account. That 
is in the $Z$ coupling to a fermion $f$ one should make additional 
replacement:
\be\label{vsfshift}
v_S^f\rightarrow v_S^f-2\Delta(\sin^2\theta_W)Q^f=v_S^f+2\kappa Q^f\Delta\varrho_M
\ee
with $\Delta(\sin^2\theta_W)$ given in (\ref{swshift}). This contribution to 
the shift $\delta\mathcal{O}$ is represented by the first term (\ref{ALT}).
\item
Denoting the result for a given observable calculated in this manner by 
$\mathcal{O}(\xi)$ we can simply calculate the shift $\delta\mathcal{O}$
due to NP by subtracting the pure SM contributions at tree level:
\be\label{Onum}
\delta\mathcal{O}(\xi)=\mathcal{O}(\xi)-\mathcal{O}(0).
\ee
As the mixing effects are very small the higher order terms generated 
by this numerical procedure are tiny and consequently one would obtain 
quite generally  very similar results by using the linearized form in (\ref{ALT}) as done in 
\cite{Altarelli:1996pr,Chiappetta:1996km}. Moreover as stressed in 
\cite{Altarelli:1996pr} in certain models such a linearized expression 
could not be sufficiently precise and using (\ref{Onum}) from the start 
takes such effects into account.
\item
For the SM contributions we will use the most recent values for EWPO that include electroweak radiative corrections. Even if the interference terms 
between SM and NP contributions will not include these corrections, this 
procedure is justified in view of the smallness of NP effects in 331 models 
considered by us.
\item
Comparing SM results with the data we will be able to identify the pattern 
of deviations from SM predictions  and see for which 
values of $\beta$, $\tan\bar\beta$ and fermion representations the 331 
models can improve the agreement of the theory with the  data and 
what does this imply for our analysis of flavour violating effects.
\end{itemize}
\subsection{Basic Formulae for EWPO}
The tree level formula for the partial widths is given by 
\be\label{GZff}
\Gamma(Z\to\bar ff)= \frac{G_FM_Z^3}{6\pi\sqrt{2}}\left[\frac{\cos^2\theta_W}{g^2}\right]\varrho N_c
\left[[\Delta^{f}_V(Z)]_{\rm eff}^2+[\Delta^{f}_A(Z)]_{\rm eff}^2\right]
\ee
with $N_c=3$ for quarks and $N_c=1$ for leptons. The additional overall factor 
relative to formula (2.8) in \cite{Altarelli:1996pr}  takes into account the difference in the 
definitions of the vector and axial-vector couplings summarized above. 
Defining next $\Gamma_f = \Gamma(Z\to\bar ff)$ we have 
\be
\Gamma_\ell=\Gamma_\mu, \qquad \Gamma_h=2\Gamma_c+2\Gamma_u + \Gamma_b, \qquad \Gamma_T=\Gamma_h+3(\Gamma_\mu+\Gamma_\nu).
\ee
While we do not distinguish between $\varrho$ for  $b$ quark contributions and contributions from lighter quarks as such effects are taken into account in 
the full SM contribution, we separate NP $b$ quark contribution to 
$\Gamma_h$ from the one of $d$ and $s$ as transformation properties of the 
third generation of quarks under $SU(3)_L$ are different compared to the 
the first two generations. See the Lagrangian~(63) in  \cite{Buras:2012dp}. 
This difference has to be taken into account in all observables involving 
the $b$ quark.

Of interest are then the ratios $R_f$ and the peak cross sections $\sigma_p^f$ 
defined as follows
\be
R_\ell=\frac{\Gamma_h}{\Gamma_\ell},\qquad R_b=\frac{\Gamma_b}{\Gamma_h},\qquad 
 R_c=\frac{\Gamma_c}{\Gamma_h}\, \qquad \sigma_p^f=\frac{12\pi}{M_Z^2}\frac{\Gamma_e\Gamma_f}{\Gamma_T^2}.
\ee
Note that the definition of  $R_\ell$ differs from the one of $R_c$ and $R_b$.
For the asymmetries $A_f$ we have
\be
\mathcal{A}_f= \frac{2 [\Delta^{f}_A(Z)]_{\rm eff} [\Delta^{f}_V(Z)]_{\rm eff}}
               {[\Delta^{f}_V(Z)]_{\rm eff}^2+[\Delta^{f}_A(Z)]_{\rm eff}^2}\,.
\ee
and for the forward-backward asymmetry
\be
\mathcal{A}^f_{\rm FB}=\frac{3}{4} \mathcal{A}_e \mathcal{A}_f\,.
\ee

We would like to warn the reader that similar to  \cite{Altarelli:1996pr} our 
 vector and axial-vector couplings for $Z$ and $Z^\prime$ are defined in terms 
of left-handed and right-handed ones as follows:
\be\label{VA}
\Delta_V=\Delta_R+\Delta_L,\qquad \Delta_A=\Delta_R-\Delta_L\,.
\ee
Consequently the axial-vector couplings differ by sign from the ones used 
in PDG. This implies that also asymmetries $\mathcal{A}_f$ differ by sign from PDG and in Table~\ref{tab:EWPfit} when quoting PDG values we adjusted their 
definition to our.

From these formulae it is straightforward to derive general analytical formulae for the coefficients $A_{\mathcal{O}}$ and $B_{\mathcal{O}}$ in (\ref{ALT}), valid in any $Z^\prime$ model.
We first find in the case of the partial width $\Gamma_f$
\be\label{Gf1}
A_{\Gamma_f}=4\kappa Q^f\frac{v_S^f}{(v_S^f)^2+(a_S^f)^2}+1\,,\qquad 
B_{\Gamma_f}=2 \frac{v_N^ fv_S^f+a_N^ fa_S^f}{(v_S^f)^2+(a_S^f)^2},
\ee
where the second term in $A_{\Gamma_f}$ is the universal correction from 
the shift in $\varrho$ in the overall factor in (\ref{GZff}). 

In the case of the asymmetry $\mathcal{A}_f$  the corresponding expressions read
\be\label{Af1}
A_{\mathcal{A}_f}=2\kappa \frac{Q^f}{v_S^f}\frac{(a_S^f)^2-(v_S^f)^2}{(v_S^f)^2+(a_S^f)^2}\,,\qquad
B_{\mathcal{A}_f}=\frac{v_N^f}{v_S^f}+\frac{a_N^f}{a_S^f}-2\frac{v_N^ fv_S^f+a_N^ fa_S^f}{(v_S^f)^2+(a_S^f)^2}\,.
\ee
 Our results in (\ref{Gf1}) and (\ref{Af1} agree with the ones
one would derive from formulae  (6.5)-(6.6) and  (7.4)-(7.7) 
in \cite{Altarelli:1990dt} by dividing them by $\Gamma_f$ and 
$\mathcal{A}_f$, respectively.

 For $R_b$ and $R_c$ we find respectively
\be\label{ARb}
A_{R_b}=A_{\Gamma_b}-\left(2\frac{\Gamma_c}{\Gamma_h}A_{\Gamma_c}+3\frac{\Gamma_b}{\Gamma_h}A_{\Gamma_b}\right), \qquad 
B_{R_b}=B_{\Gamma_b}-\left(2\frac{\Gamma_c}{\Gamma_h}B_{\Gamma_c}+3\frac{\Gamma_b}{\Gamma_h}B_{\Gamma_b}\right)
\ee
and
\be\label{ARc}
A_{R_c}=A_{R_b}+A_{\Gamma_c}-A_{\Gamma_b},
\qquad B_{R_c}=B_{R_b}+B_{\Gamma_c}-B_{\Gamma_b}
\ee
In these formulae $\Gamma_f$ and $\Gamma_h$ are just tree-level SM contributions.

We observe that indeed the coefficients $A_{\mathcal{O}}$ depend only on SM couplings 
and the first term in (\ref{ALT}) feels the presence of $Z^\prime$ only through 
$\sin\xi$, while the second one has additional dependence on the diagonal 
$Z^\prime$ couplings. While these formulae allow an easy comparison with the 
analyses in \cite{Altarelli:1990dt,Altarelli:1996pr,Chiappetta:1996km}, in 
 numerical calculations it is easier to use directly (\ref{Onum}).

\subsection{Numerical Results}

\begin{table}[!tb]
{\renewcommand{\arraystretch}{1.3}
\begin{center}
\begin{tabular}{|c||c|c|c|c|}
\hline
$\beta$ & $-\frac{2}{\sqrt{3}}$ & $-\frac{1}{\sqrt{3}}$ & $\frac{1}{\sqrt{3}}$ & $\frac{2}{\sqrt{3}}$ \\
\hline
$\delta\Gamma_Z$ $[10^{-3}]$ & $0.132(2.582) $ & $ 0.0269(1.308)  $ & $ 0.0152( 0.578) $& $ 0.0949(0.422) $\\ 
$\delta\sigma_h$  $\text{[nbarn}\times10^{-3}]$& $23.48(58.90)  $ & $ 7.678(30.77) $ & $-7.883(16.04)  $ & $ -24.69(13.00) $ \\
$\delta R_\ell$  $[10^{-3}]$& $-5.250(-7.828) $ & $ -2.249(-6.143) $ & $  3.395( -5.474) $ & $ 12.23(-5.339) $\\
$\delta\mathcal{A}^\ell_{\rm FB}$  $[10^{-3}]$& $0.472(1.838) $ & $0.0924(0.705) $ & $ 0.0291(0.104)  $ & $ 0.274(-0.0186)  $\\
$\delta\mathcal{A}_\ell$  $[10^{-3}]$& $-2.084(-7.964) $ & $ -0.410(-3.106) $ & $-0.129( -0.463) $ & $-1.215(0.0826)  $\\
$\delta\mathcal{A}_c$  $[10^{-3}]$& $-0.915(-3.493) $    & $  -0.180(-1.364)$  & $ -0.0569( -0.203) $  & $ -0.533( 0.0363) $ \\
$\delta\mathcal{A}_b$ $[10^{-3}]$& $ -0.168(-0.643) $ & $ -0.0332(-0.251)  $ & $  -0.0105( -0.0374) $ & $  -0.0983(0.00668) $\\
$\delta\mathcal{A}^c_{\rm FB}$  $[10^{-3}]$& 1.149(4.408)   & 0.226(1.714)  &  0.0713( 0.255)  &  0.670(-0.0455)  \\
$\delta\mathcal{A}^b_{\rm FB}$  $[10^{-3}]$& $ 1.482(5.665)  $ & $  0.292(2.209) $ & $ 0.0920(0.329) $ & $ 0.864(-0.0587)  $\\
$\delta R_c$  $[10^{-3}]$& $ 0.0979(0.269)  $ & $ 0.0296( 0.132) $ & $  -0.0255( 0.0584) $ & $  -0.0727(0.0432) $ \\
$\delta R_b$  $[10^{-3}]$& $ 0.0950(0.227) $ & $  0.0320(0.123) $ & $  -0.0348( 0.0683) $ & $ -0.112( 0.0569) $ \\
$\Omega^{331}$ & $17.70 (55.25) $ & $15.45(21.32) $ & $16.39( 15.20) $ & $19.42( 15.02) $ \\
$\Omega^{331}$(LEP) & $18.29(53.59) $ & $13.95(22.59) $ & $14.42(13.78) $ & $ 19.08(12.66)$ \\
$\delta\sin^2\theta^\ell_\text{eff}$ $[10^{-3}]$& $-0.2650 (-1.0135)$ & $-0.0522(-0.3950)$ & $-0.0165(-0.0589)$ & $-0.1545(0.01050)$\\
\hline
\end{tabular}
\end{center}}
\caption{\it Values of the shifts in EWPO for different $\beta$ and $\tan\bar\beta=1(5)$ in scenario $F_1$ for fermion representations. We also give the values of $\Omega^{331}$ defined in (\ref{OMEGA}) and 
$\delta\sin^2\theta^\ell_\text{eff}$  from~(\ref{equ:sineff}).
\label{tab:EWPO1}}~\\[-2mm]\hrule
\end{table}

\begin{table}[!tb]
{\renewcommand{\arraystretch}{1.3}
\begin{center}
\begin{tabular}{|c||c|c|c|c|}
\hline
$\beta$ & $-\frac{2}{\sqrt{3}}$ & $-\frac{1}{\sqrt{3}}$ & $\frac{1}{\sqrt{3}}$ & $\frac{2}{\sqrt{3}}$ \\
\hline
$\delta\Gamma_Z$ $[10^{-3}]$ & $1.057(4.924) $ & $ 0.176(1.917)  $ & $ 0.165(0.268) $& $ 1.021(-0.0730)  $\\ 
$\delta\sigma_h$  $\text{[nbarn}\times10^{-3}]$& $24.18(60.32)  $ & $ 7.795(31.20) $ & $ -7.764(15.80) $ & $ -23.96(12.62) $ \\
$\delta R_\ell$  $[10^{-3}]$& $ -9.294(-17.88) $ & $  -2.906(-8.787) $ & $  2.737(-4.119) $ & $ 8.165( -3.168)$\\
$\delta\mathcal{A}^\ell_{\rm FB}$  $[10^{-3}]$& $ 0.0375(0.706) $ & $ 0.0227(0.417) $ & $  -0.0405( 0.249) $ & $-0.158( 0.213)  $\\
$\delta\mathcal{A}_\ell$  $[10^{-3}]$& $-0.167(-3.108) $ & $  -0.101(-1.845) $ & $0.180(-1.104) $ & $ 0.703(-0.943) $\\
$\delta\mathcal{A}_c$  $[10^{-3}]$& $ -0.0733( -1.365) $    & $ -0.0442(-0.811) $  & $0.0792(-0.485) $  & $0.309( -0.414)  $ \\
$\delta\mathcal{A}_b$ $[10^{-3}]$& $-0.0135(-0.251) $ & $ -0.00814( -0.149)  $ & $0.0146(-0.0893)  $ & $ 0.0568( -0.0763)  $\\
$\delta\mathcal{A}^c_{\rm FB}$  $[10^{-3}]$&  0.0919(1.715)  &  0.0554( 1.018)  &  $-0.0994(0.609)$  &   $-0.387( 0.520)$  \\
$\delta\mathcal{A}^b_{\rm FB}$  $[10^{-3}]$& $ 0.119( 2.210)  $ & $ 0.0715( 1.312) $ & $ -0.128(0.785) $ & $ -0.500( 0.671)  $\\
$\delta R_c$  $[10^{-3}]$& $  0.0832(0.232)  $ & $  0.0273( 0.122) $ & $ -0.0279( 0.0632) $ & $ -0.0873 ( 0.0511) $ \\
$\delta R_b$  $[10^{-3}]$& $ 0.105( 0.252) $ & $0.0336( 0.129) $ & $  -0.0332(0.0650) $ & $-0.102 (0.0517) $ \\
$\Omega^{331}$ & $14.96(26.29) $ & $15.28(17.77) $ & $16.31( 15.90) $ & $18.31(15.76) $ \\
$\Omega^{331}$(LEP) &13.04(27.57) &13.26(18.14) & 13.77(15.41) &14.74(15.06) \\
$\delta\sin^2\theta^\ell_\text{eff}$ $[10^{-3}]$& $ -0.0212(-0.3953)$ & $ -0.0128(-0.2347)$ & $0.0229(-0.1404) $ & $0.0893(-0.1200) $\\
\hline
\end{tabular}
\end{center}}
\caption{\it Values of the shifts in EWPO for different $\beta$ and $\tan\bar\beta=1(5)$ in scenario $F_2$ for fermion representations. We also give the values of $\Omega^{331}$ defined in (\ref{OMEGA}) and 
$\delta\sin^2\theta^\ell_\text{eff}$  from~(\ref{equ:sineff}).
\label{tab:EWPO2}}~\\[-2mm]\hrule
\end{table}

\begin{table}[!tb]
{\renewcommand{\arraystretch}{1.3}
\begin{center}
\begin{tabular}{|c||c|c|c|c|}
\hline
$\beta$ & $-\frac{2}{\sqrt{3}}$ & $-\frac{1}{\sqrt{3}}$ & $\frac{1}{\sqrt{3}}$ & $\frac{2}{\sqrt{3}}$ \\
\hline
$\delta\Gamma_Z$ $[10^{-3}]$ & $0.403(-0.092) $ & $ 0.553 (0.244)$ & $ 1.260 (1.869)$& $2.485 (4.831) $\\ 
$\delta\sigma_h$ $\text{[nbarn}\times10^{-3}]$& $-12.76(-13.15)  $ & $-16.17 (-16.41) $ & $-32.60 (-32.14) $ & $-63.39 (-61.75) $ \\
$\delta R_\ell$  $[10^{-3}]$& $4.003(6.181) $ & $6.208 (7.568) $ & $ 16.857 (14.193)$ & $36.619 (26.418) $\\
$\delta\mathcal{A}^\ell_{\rm FB}$  $[10^{-3}]$& $-0.123(0.108) $ & $-0.027 (0.118) $ & $0.445 (0.158) $ & $1.322 (0.205) $\\
$\delta\mathcal{A}_\ell$  $[10^{-3}]$& $0.548(-0.478) $ & $0.119 (-0.523) $ & $-1.964 (-0.703) $ & $ -5.771 (-0.908)$\\
$\delta\mathcal{A}_c$  $[10^{-3}]$& $0.241(-0.210)$   &  $0.052 (-0.230)$ & $-0.863 (-0.309)$  & $-2.532 (-0.399) $ \\
$\delta\mathcal{A}_b$ $[10^{-3}]$& $ 0.044(-0.039) $ & $0.010 (-0.042) $ & $ -0.159 (-0.057)$ & $-0.466 (-0.073) $\\
$\delta\mathcal{A}^c_{\rm FB}$  $[10^{-3}]$& $-0.302(0.264)$  &$ -0.065 (0.288) $  & $1.084 (0.387)  $  &  $3.190 (0.500) $  \\
$\delta\mathcal{A}^b_{\rm FB}$  $[10^{-3}]$& $-0.389(0.340) $ & $-0.084 (0.372) $ & $1.397 ( 0.500)$ & $4.105 (0.645) $\\
$\delta R_c$  $[10^{-3}]$& $-0.048(-0.040) $ & $-0.056 (-0.051) $ & $-0.093 (-0.102)  $ & $ -0.163 (-0201)$ \\
$\delta R_b$  $[10^{-3}]$& $-0.054(-0.059) $ & $ -0.070 (-0.073)$ & $ -0.149 (-0.143)$ & $ -0.297 (-0.272)$ \\
$\Omega^{331}$ & $ 16.84(17.12) $ & $17.20(17.53) $ & $22.55( 20.42) $ & $48.04( 29.70) $ \\
$\Omega^{331}$(LEP) & 13.59(15.72) & 14.78(16.20) & 23.03(19.37) &48.97(28.94) \\
$\delta\sin^2\theta^\ell_\text{eff}$ $[10^{-3}]$& $0.0697(-0.0608) $ & $0.0151 (-0.0665)$ & $-0.2498(-0.0894) $ & $-0.7342(-0.1154) $\\
\hline
\end{tabular}
\end{center}}
\caption{\it Values of the shifts in EWPO for different $\beta$ and $\tan\bar\beta=0.2$ in scenario $F_1(F_2)$ for fermion representations. We also give the values of $\Omega^{331}$ defined in (\ref{OMEGA}) and 
$\delta\sin^2\theta^\ell_\text{eff}$  from~(\ref{equ:sineff}).
\label{tab:EWPO3}}~\\[-2mm]\hrule
\end{table}

\begin{table}[!tb]
{\renewcommand{\arraystretch}{1.3}
\begin{center}
\begin{tabular}{|c||c|c|c|}
\hline
Quantity & Input Data & SMfit & Pull  \\
\hline
$\Gamma_Z$ & $2.4952(23)$ & $2.4954(14)$ & $0.09$\\ 
$\sigma_h~$[nbarn] & $41.540(37)$ & $ 41.479(14)$ & $-1.65$  \\
$R_\ell$ & $20.767(25)$ & $20.740(17)$ & $-1.08$ \\
$\mathcal{A}^\ell_{\rm FB}$& $0.0171(10)$ & $0.01627(2)$ & $-0.83$ \\
$\mathcal{A}_\ell({\rm LEP})$ &  $-0.1465(33)$ & $-0.1472(7)$ & $-0.2$  \\
$\mathcal{A}_\ell({\rm SLD})$ &  $-0.1513(21)$ & $-0.1472(7)$ & $1.95$  \\
$\mathcal{A}_\ell$ &  $-0.1499(18)$ & $-0.1472(7)$ & $1.50$ \\
$\sin^2\theta^l_\text{eff}$ & $0.2324(12)$ & $0.23148(10)$ & $-0.7$ \\
$\mathcal{A}_c$& $-0.670(27)$ & $-0.6679(3)$ & $0.07$  \\
$\mathcal{A}_b$& $-0.923(20)$ & $-0.93464(5)$ & $-0.58$ \\
$\mathcal{A}^c_{\rm FB}$ & $0.0707(35)$& $0.0738(4)$ & $0.88$  \\
$\mathcal{A}^b_{\rm FB}$ & $0.0992(16)$ & $0.1032(5)$ & $2.5$ \\
$R_c$ & $0.1721(30)$ & $0.17223(6)$ & $0.04$  \\
$R_b$ & $0.21629(66)$ & $0.21548(5)$ & $-1.23$  \\
\hline
\end{tabular}
\end{center}}
\caption{\it Input Data and SM fit for 
various  EWPO and the pull values. Update of \cite{Baak:2012kk}. 
\label{tab:EWPfit}}~\\[-2mm]\hrule
\end{table}

In Tables~\ref{tab:EWPO1}-\ref{tab:EWPO3} we list the shifts in a number of 
observables as functions of $\beta$, $\tan\bar\beta$ for the fermion representations $F_1$ and $F_2$. In Table~\ref{tab:EWPfit} we summarize SM predictions for 
these observables, the corresponding data and the pulls as presented after 
Higgs discovery in \cite{Baak:2012kk}. 

In what follows it will be useful to denote by MI, with a given I=1,..24,
 a particular 331 model in which $\beta$, $\tan\bar\beta$ and fermion representations are fixed. The index I numbers the column in  Tables~\ref{tab:EWPO1}-\ref{tab:EWPO3} corresponding to a given model in order of its appearance.
Thus we deal with 24 models { that we specify in Table \ref{tab:331models} to make their identification easier}. 
For instance M5 denotes the model with $\beta=1/\sqrt{3}$, $\tan\bar\beta=1$ and $F_1$. 
\begin{table}[!tb]
{\renewcommand{\arraystretch}{1.3}
\begin{center}
\begin{tabular}{|c||c|c|c||c||c|c|c||c||c|c|c|}
\hline
MI  &     {\rm scen.} &  $\beta$ & $\tan {\bar \beta}$ & MI  & {\rm scen.} &  $\beta$ & $\tan {\bar \beta}$ &MI  & {\rm scen.} &  $\beta$ & $\tan {\bar \beta}$\\
\hline
M1 & $F_1$ & $-2/\sqrt{3}$ & 1 & M9 & $F_2$ & $-2/\sqrt{3}$ & 1 & M17 & $F_1$ & $-2/\sqrt{3}$ & 0.2
\\
M2 & $F_1$ & $-2/\sqrt{3}$ & 5 & M10 & $F_2$ & $-2/\sqrt{3}$ & 5 & M18 & $F_2$ & $-2/\sqrt{3}$ & 0.2
\\
M3 & $F_1$ & $-1/\sqrt{3}$ & 1 & M11 & $F_2$ & $-1/\sqrt{3}$ & 1 & M19 & $F_1$ & $-1/\sqrt{3}$ & 0.2
\\
M4 & $F_1$ & $-1/\sqrt{3}$ & 5 & M12 & $F_2$ & $-1/\sqrt{3}$ & 5 & M20 & $F_2$ & $-1/\sqrt{3}$ & 0.2
\\
M5 & $F_1$ & $1/\sqrt{3}$ & 1 & M13 & $F_2$ & $1/\sqrt{3}$ & 1 & M21 & $F_1$ & $1/\sqrt{3}$ & 0.2
\\
M6 & $F_1$ & $1/\sqrt{3}$ & 5 & M14 & $F_2$ & $1/\sqrt{3}$ & 5 & M22 & $F_2$ & $1/\sqrt{3}$ & 0.2
\\
M7 & $F_1$ & $2/\sqrt{3}$ & 1 & M15 & $F_2$ & $2/\sqrt{3}$ & 1 & M23 & $F_1$ & $2/\sqrt{3}$ & 0.2
\\
M8 & $F_1$ & $2/\sqrt{3}$ & 5 & M16 & $F_2$ & $2/\sqrt{3}$ & 5 & M24 & $F_2$ & $2/\sqrt{3}$ & 0.2
\\
\hline
\end{tabular}
\end{center}}
\caption{\it Definition of the various 331 models.
\label{tab:331models}}~\\[-2mm]\hrule
\end{table}

In order to judge the quality of a given model and compare it with the 
performance of the SM we define for each observable $\mathcal{O}_i$ the {\it pulls} $P_i^\text{SM}$ and $P_i^{331}$ as follows
\be\label{Pulls}
P_i^\text{SM}=\frac{\text{SMfit}_i-(\text{Input~Data})_i}{\sigma_\text{exp}^i},
\qquad P_i^{331}=\frac{\text{SMfit}_i+\delta\mathcal{O}_i-(\text{Input~Data})_i}{\sigma_\text{exp}^i}.
\ee
The pulls $P_i^\text{SM}$ are the usual ones and their values are given in the 
last column in   Table~\ref{tab:EWPfit}. In principle in order to calculate 
such pulls for every 331 model MI considered by us we would have to 
repeat the fit of \cite{Baak:2012kk} for all models including also other observables that are sensitive to new charged gauge bosons. Such an analysis is clearly 
beyond the scope of our paper. As $\mathcal{O}_i$ are fixed in a given MI, 
that is do not introduce new parameters, we expect that such a simplified 
procedure should give us a correct, even if rough, picture of what is going on.

In order to identify favourite 331 models, as far as electroweak observables are 
concerned, we define the measures
\be\label{OMEGA}
\Omega^\text{SM}=\sum_i \left(P_i^\text{SM}\right)^2= { 15.72~ (13.51)}, \qquad
\Omega^{331}=\sum_i \left(P_i^{331}\right)^2\, .
\ee
The first SM value is based on Table~\ref{tab:EWPfit} using the average of LEP 
and SLD values for $\mathcal{A}_\ell$ while the one in the parentheses 
is obtained by using as curiosity only the LEP value.
The values of $\Omega^{331}$ 
for 331 models are given in Tables~\ref{tab:EWPO1}-\ref{tab:EWPO3}. The models with smallest $\Omega^{331}$  are favoured, while the ones with 
with largest $\Omega^{331}$ disfavoured.

Inspecting these results we make first the following observations when 
the average of LEP and SLD values for $\mathcal{A}_\ell$ is used.
\begin{itemize}
\item
All models give small contributions to $\mathcal{A}_c$ and $R_c$ and consequently agree with the data.
\item
All models give rather small contributions to $\mathcal{A}_b$ and 
consequently cannot remove the small pull present in the SM. Moreover, 
M21 and M23 soften the agreement of the theory with data.
\item
The sizable discrepancy of the SM with the data on $R_b$ cannot be removed 
in any model. It can be softened by $(0.3-0.4)\sigma$ in M2 and M10 and visibly 
increased in models M21-M24.
\end{itemize}

We concentrate then our discussion on the remaining observables, where 
NP effects turn out to be larger.  We find
\begin{itemize}
\item
In most models NP effects in $\Gamma_Z$ are small in agreement with data. 
Deviations larger than $1\sigma$ are only found in M2, M10 and M23.
\item
In the case of $\sigma^h$ the agreement with data is significantly improved 
in the case of M1, M2, M4, M9, M10, M12 but worsened visibly in M7, M15 and 
all models with $\tan\bar\beta=0.2$, that is M17-M24. 
\item
Interestingly in the case of $R_\ell$ all the models with $\tan\bar\beta=0.2$ 
improve the agreement of the theory with data and this also applies to M7 
and  M15, while the remaining models slightly worsen the agreement. 
\item 
On the contrary in the case of $\mathcal{A}^\ell_{\rm FB}$ and $\mathcal{A}_\ell$ 
the models with $\tan\bar\beta=1.0$ and $\tan\bar\beta=5.0$ are performing 
much better than the ones with $\tan\bar\beta=0.2$. Moreover they 
perform better than the SM. However, this depends, 
as discussed below, whether one takes into account the SLD value for $\mathcal{A}_\ell$ or not.
\item
Concerning $\delta\mathcal{A}^c_{\rm FB}$ and  $\delta\mathcal{A}^b_{\rm FB}$ 
the favourite models listed below introduce only a very small shift to the 
SM values not improving the status of the theory, while several models, 
in particular M2, M4, M7, M10, M12, M21 and M23 worsen the agreement with 
the data. This is also the case of two favoured models below, M14 and M16 
but they compensate it through better results for $\sigma^h$ than obtained 
within the SM.
\end{itemize}

The final verdict is given by the values of $\Omega^{331}$  
in Tables~\ref{tab:EWPO1}-\ref{tab:EWPO3}. We observe that seven  331 models 
 have  { $\Omega^{331}< 16.0$.} These are in the order of increasing 
 $\Omega^{331}$
\be\label{favoured}
{\rm M9}, \quad {\rm M8},\quad {\rm M6}, \quad {\rm M11}, \quad {\rm M3}, \quad {\rm M16}, \quad {\rm M14}, \qquad {(\rm favoured)}
\ee
with the first five performing better than the SM while the last two 
having basically the same $\Omega^{331}$.
The models with {\it odd} index I correspond to $\tan\bar\beta=1.0$ and 
the ones with {\it even} one to  $\tan\bar\beta=5.0$.   We list the pulls 
$P_i^{331}$ in these seven models in Table~\ref{tab:331pulls}.

\begin{table}[!tb]
{\renewcommand{\arraystretch}{1.3}
\begin{center}
\begin{tabular}{|c||c|c|c|c|c|c|c|}
\hline
Pull  &    \begin{tabular}{l} \footnotesize{M3}\\ \tiny{
$\beta=-\frac{1}{\sqrt{3}}$, \,F1}\\ \tiny{$\tan \bar \beta =1$ }\end{tabular}&   
 \begin{tabular}{l} \footnotesize{M6} \\\tiny{
$\beta=\frac{1}{\sqrt{3}}$, \,F1}\\ \tiny{$\tan \bar \beta =5$ }\end{tabular}&   
 \begin{tabular}{l}  \footnotesize{M8}\\ \tiny{
$\beta=\frac{2}{\sqrt{3}}$, \,F1}\\ \tiny{$\tan \bar \beta =5$ }\end{tabular}&  
 \begin{tabular}{l} \footnotesize{M9} \\\tiny{
$\beta=-\frac{2}{\sqrt{3}}$, \,F2}\\\tiny{$\tan \bar \beta =1$ }\end{tabular}&   
 \begin{tabular}{l} \footnotesize{M11 } \\ \tiny{
$\beta=-\frac{1}{\sqrt{3}}$, \,F2}\\ \tiny{$\tan \bar \beta =1$ }\end{tabular}&   
 \begin{tabular}{l} \footnotesize{M14} \\ \tiny{
$\beta=\frac{1}{\sqrt{3}}$, \,F2}\\ \tiny{$\tan \bar \beta =5$ }\end{tabular}& 
\begin{tabular}{l}  \footnotesize{M16} \\ \tiny{
$\beta=\frac{2}{\sqrt{3}}$, \,F2}\\ \tiny{$\tan \bar \beta =5$ }\end{tabular} \\
\hline
$P_{\Gamma_Z}$ & 0.099 & 0.338 & 0.271 & 0.546 & 0.164 & 0.204 & 0.055
\\
$P_{\sigma_h}$  &  $-1.441$ & $-1.215$ & $-1.297$ & $-0.995$ & $-1.438$ & $-1.222$ & $-1.308$
\\
$P_{ R_\ell}$  & $-1.170$ & $-1.299$ & $-1.294$ &$-1.452$ & $-1.196$ & $-1.245$ & $-1.207$
\\
$P_{\mathcal{A}^\ell_{\rm FB}}$ & $-0.737$ & $-0.726$ &$-0.849$ & $-0.792$ & $-0.807$ & $-0.581$ & $-0.617$
 \\
$P_{\mathcal{A}_\ell}$ & 1.216 & 1.187 & 1.490 & 1.352 & 1.389 &0.831 & 0.920
\\
$P_{\mathcal{A}_c}$  & 0.067 & 0.067 & 0.075 &0.071 & 0.072 & 0.056 &0.059
\\
$P_{\mathcal{A}_b}$& $-0.584$ & $-0.584$ &$-0.582$ & $-0.583$ & $-0.582$ & $-0.586$ & $-0.586$
\\
$P_{\mathcal{A}^c_{\rm FB}}$  & 0.979 & 0.987 & 0.901& 0.941 & 0.930 & 1.088 & 1.063
\\
$P_{\mathcal{A}^b_{\rm FB}}$  & 2.682 & 2.706 & 2.463 & 2.574 & 2.545 & 2.991 & 2.919
\\
$P_{R_c}$  &0.053 & 0.063 &  0.058 & 0.071 & 0.052 & 0.064 & 0.060
\\
$P_{R_b}$ & $-1.179$ & $-1.124$ & $-1.141$ &$-1.069$ & $-1.176$ & $-1.129$ & $-1.149$
\\
\hline
\end{tabular}
\end{center}}
\caption{\it Pulls for the seven selected 331 models.
\label{tab:331pulls}}~\\[-2mm]\hrule
\end{table}

As seen in this table all favoured models improve the agreement of the 
theory with data on $\sigma^h$. This is in particular the case of M9 and this 
fact is primarily responsible why M9 wins the competition as it does reasonably 
well for other observables.
We also observe that all favoured models, except M8, improve the agreement of theory with 
data on  $\mathcal{A}^\ell_{\rm FB}$ and $\mathcal{A}_\ell$. However the models 
M14 and M16 which perform best in this respect are not the top models 
as they perform worse than the first ones on other observables 
which basically do not provide any 
improvements on these asymmetries. The reason is that M14 and M16 experience 
difficulties with  $\delta\mathcal{A}^c_{\rm FB}$ and  $\delta\mathcal{A}^b_{\rm FB}$, as stated above, making the agreement of the theory with data to be 
worse than in the SM. 

Yet, as discussed below, our analysis confirms the general findings of 
Richard \cite{Richard:2013xfa} that in 331 models the departure of the 
data on  $\mathcal{A}^\ell_{\rm FB}$ and $\mathcal{A}_\ell$ from their SM values is 
correlated within 331 models with  the $B_d\to K^*\mu^+\mu^-$ anomaly, 
even if the model M2 which he studied is basically excluded by 
other electroweak data when correct expression for $\sin\xi$ is used.
This is  in particular the case of $\Gamma_Z$ and $\mathcal{A}^b_{\rm FB}$.
 Otherwise this model is interesting as it removes the discrepancy of 
the SM with the data on $\sigma^h$.

An important result of our analysis is the weak 
performance of models with $\tan\bar\beta=0.2$, although the models M17-M20 cannot be excluded. The models which definitely have difficulties with electroweak precision data are 
\be\label{disfavoured}
{\rm M2}, \quad {\rm M4},\quad {\rm M7}, \quad {\rm M10}, \quad {\rm M21}, \quad{\rm M22}, \quad {\rm M23}, \quad {\rm M24}, \qquad {(\rm disfavoured)}
\ee

Looking at these results we conclude that from the point of view of electroweak 
precision tests the following combinations of the values of $\beta$, $\tan\bar\beta$ and fermion representations are favoured.
\begin{itemize}
\item
Three models with $\beta=-2/\sqrt{3}$ and $\beta=-1/\sqrt{3}$ for 
$\tan\beta=1.0$. For $F_1$ this is M3 with $\beta=-1/\sqrt{3}$ and 
for $F_2$ M9 and M11 for $\beta=-2/\sqrt{3}$ and $\beta=-1/\sqrt{3}$, respectively.
\item
Four models with $\beta=1/\sqrt{3}$ and $\beta=2/\sqrt{3}$ for $\tan\beta=5.0$.
For $F_1$ these are M6 and M8 for $\beta=1/\sqrt{3}$ and $\beta=2/\sqrt{3}$, 
respectively and analogously M14 and M16 in the case for $F_2$.
\end{itemize}
It should be emphasized that all these seven models pass the present 
electroweak tests for $M_{Z^\prime}=3\tev$ as well as the SM or in five cases even
better than it. Even if the models with $\beta=\pm 2/\sqrt{3}$ perform 
slightly better than the ones with $\beta=\pm 1/\sqrt{3}$ it is not 
possible on the basis of $\Omega^{331}$ alone to identify the winner among them.
On the other hand one day this should be possible  on the basis of the Table~\ref{tab:331pulls} 
once various questions related to measurements at LEP and SLD will be clarified.
 As demonstrated below flavour physics can offer definite help in this context.

In the latter context it is interesting to observe that if the LEP result 
for $\mathcal{A}_\ell$ would be the true one five models on the list of 
favourites in (\ref{favoured}) would remain 
\be\label{favouredLEP}
{\rm M8}, \quad {\rm M9},\quad {\rm M11}, \quad {\rm M17}, \quad {\rm M13}, \quad {\rm M6}, \quad {\rm M3}, \qquad {(\rm LEP~ favoured)}
\ee
with the first three  performing better than the SM. M14 and M16 are not 
present anymore on this list because they favoured SLD result. Instead M13 
and in particular M17 with $\tan\bar\beta=0.2$ are among favourites. As 
we will discuss below M17 has a unique property among the favourites as 
far as flavour physics is concerned.

\boldmath
\subsection{The issue of $\sin^2\theta^\ell_\text{eff}$}
\unboldmath
In testing the SM one can define  $\sin^2\theta^\ell_\text{eff}$  by 
using  SM tree-level expression for $\mathcal{A}_\ell$. This parameter
is most precisely extracted from the data on $\mathcal{A}_\ell$  and 
$\mathcal{A}^b_{\rm FB}$ but also from $\mathcal{A}^\ell_{\rm FB}$. 
Unfortunately the determinations of $\sin^2\theta^\ell_\text{eff}$ from 
these observables are really not in agreement with each other. On one hand 
in the case of $\mathcal{A}_\ell$ we have \cite{Tenchini:2008zz,Richard:2013xfa} \be
\sin^2\theta^\ell_\text{eff}(\text{SLD})=0.23098 (26),\qquad 
\sin^2\theta^\ell_\text{eff}(\text{LEP})=0.23159(41)
\ee
and from forward-backward asymmetries  $\mathcal{A}^\ell_{\rm FB}$ and 
$\mathcal{A}^b_{\rm FB}$ one finds respectively
\be
\sin^2\theta^\ell_\text{eff}=0.23099 (53),\qquad 
\sin^2\theta^\ell_\text{eff} =0.23221(29).
\ee

This implies roughly $3\sigma$ discrepancy between the two most precise 
determinations. The resulting values from all data as given in 
Table~\ref{tab:EWPfit} are 
 \be\label{EXPSM}
\sin^2\theta^\ell_\text{eff}(\text{EXP})=0.2324 (12),\qquad 
\sin^2\theta^\ell_\text{eff} (\text{SM}) =0.23148(10).
\ee
Consequently there is some preference for the positive shift in 
$\sin^2\theta^\ell_\text{eff}$ relative to the best SM value.

Until now we did not look at $\sin^2\theta^\ell_\text{eff}$ in 331 models and 
calculated $\mathcal{A}^\ell_{\rm FB}$, $\mathcal{A}_\ell$, $\mathcal{A}^b_{\rm FB}$  and other observables directly to
judge the quality of a given model on the basis of them. But it 
is instructive to calculate the shift of $\sin^2\theta^\ell_\text{eff}$ 
due to NP contributions for 24 models 
considered by us using the operative definition \cite{Tenchini:2008zz}
\be\label{equ:sineff}
\delta\sin^2\theta^\ell_\text{eff}=\frac{1}{4}\left(1+\frac{[\Delta^{\ell}_V(Z)]_{\rm eff}}{[\Delta^{\ell}_A(Z)]_{\rm eff}}\right) -\sin^2\theta_W, 
\ee
where in the effective vector couplings the shift (\ref{vsfshift}) has to be 
included. An extensive discussion of $\sin^2\theta^\ell_\text{eff}$ can 
be found in \cite{Tenchini:2008zz} where further references can be found. 
See also \cite{Richard:2013xfa}.  In writing (\ref{equ:sineff}) we 
have adjusted the sign in the formula (8.3) in \cite{Tenchini:2008zz} to our 
definition of the axial-vector coupling.

The shift in $\delta\sin^2\theta^\ell_\text{eff}$ in 331 models comes first
from the shift of $\sin^2\theta_W$ which as seen in 
(\ref{swshift}) is negative in 331 models and in any $Z^\prime$ model. 
But in addition to this shift, which comes from $\Delta\varrho_M$ and 
affects only vector couplings, both vector and axial-vector couplings receive 
modifications from the mixing with $Z^\prime$, that is the shifts in the 
couplings  proportional to $\sin\xi$ and involving $Z^\prime$ couplings. 

The result of this exercise is given in the last rows in 
 Tables~\ref{tab:EWPO1}-\ref{tab:EWPO3}. The striking feature is that out of 24 
models 19 give a negative shift of $\sin^2\theta^\ell_\text{eff}$, while only 
7 a positive one. These are M8, M13, M15, M17 and M19. 

It is not surprising that  M8, M13 and M17 perform here so well as they 
are on the list of LEP favourites 
in (\ref{favouredLEP}). What is remarkable that M8 is the only among our 
favourite models in  (\ref{favoured}) that gives a positive shift of 
 $\sin^2\theta^\ell_\text{eff}$. But this shift being $1\times 10^{-5}$ is totally 
negligible.  Yet, this model similar to all models with positive 
shift is fully compatible with the experimental value in (\ref{EXPSM}). 

We also note that models M14 and M16 on our list of favourites give
 \be\label{M14M16}
\sin^2\theta^\ell_\text{eff}(\text{M14})=0.23134,\qquad 
\sin^2\theta^\ell_\text{eff} (\text{M16}) =0.23136, 
\ee
where we have added the corresponding shifts to the SM value in (\ref{EXPSM}).
These values are within $1.5\sigma$ from the central value of $\sin^2\theta^\ell_\text{eff}$ from the SLD. 

We also confirm qualitatively the finding in \cite{Richard:2013xfa} that 
the model M2 provides a shift that results in $\sin^2\theta^\ell_\text{eff}$ 
close to the  SLD result. We find $0.23122$ and $0.23048$ for $\tan\bar\beta=1.0$ and $\tan\bar\beta=5.0$, respectively. For  $\tan\bar\beta\approx 3.0$ the 
SLD result can be well reproduced. However, this model has problems with 
other observables as we have seen above.

 While we do not think that just looking at  $\sin^2\theta^\ell_\text{eff}$ 
offers a fully transparent test of a given extension of the SM, the rather 
different values of this parameter extracted from different observables
 calls for future improved measurements of EWPO which hopefully one day 
will be possible at a future ILC. This, as stressed in  \cite{Richard:2013xfa}, 
could offer powerful tests of 331 models.

\subsection{Implications for Flavour Physics}

Our analysis of EWPO identified a group of seven models among 24 considered 
by us and the question arises whether flavour physics could distinguish 
between  them. As we will now discuss our analysis in previous sections demonstrates this rather clearly and having the plots presented there we can summarize how the seven models in question could be
 distinguished by flavour observables in the coming flavour precision era. 

Let us first summarize the main message from our analysis of EWPO which is related to the fact that the values $\tan\bar\beta=0.2$ are disfavoured:
\begin{itemize}
\item
Significant NP effects in $B$ and $K$ decays to neutrinos seem rather 
unlikely, even if the branching ratio for $\klpn$ could still be modified
by $15\%$. In turn due to the correlation with $\epe$ also NP effects
in this ratio are predicted to be small implying that $\bsi\approx 1.0$ in 
order to agree with data. On the other hand as discussed in our recent paper  
\cite{Buras:2014sba} the required precise value of this parameter depends on the values of $\vcb$ and $\vub$. 
\item 
The most interesting NP effects are found in $B\to K^*\mu^+\mu^-$ and 
$B_{s,d}\to\mu^+\mu^-$ and we will confine the discussion of the favourite 
models in (\ref{favoured}) to these decays.
\end{itemize}

The selection of favourite models and the comments just made imply 
that it is sufficient to look at Figs.~\ref{fig:ReC9vsBsmu} and 
\ref{fig:ReC9vsBsmuF2} for $F_1$ and $F_2$, respectively. The results 
for the seven favourite models can be found there with  only the results in 
the upper left panel in Fig.~\ref{fig:ReC9vsBsmu} being disfavoured. In 
the remaining seven panels in these two figures, the results for our 
favourite models are simply identified by selecting the $\tan\bar\beta=1.0$ 
line (lighter colours) in the case of $\beta < 0$ and the 
$\tan\bar\beta=5.0$ line (darker colours) in the case of $\beta >0$. The main 
implications for rare $B$ decays are then as follows:
\begin{itemize}
\item
A significant suppression of $\mathcal{B}(B_s\to\mu^+\mu^-)$ and significant 
negative shift in ${\rm Re} C_9^\text{NP}$ cannot take place simultaneously. 
This would be possible in M2 but this model belongs to disfavoured ones
by our EWPO analysis.
\item
For softening the $B_d\to K^*\mu^+\mu^-$ anomaly the most interesting is the 
model M16, that is the model with $F_2$, $\beta=2/\sqrt{3}$ and $\tan\bar\beta=5.0$. See upper right panel in Fig.~\ref{fig:ReC9vsBsmuF2}. The usual 
statements present in the literature \cite{Gauld:2013qja,Buras:2013dea,Richard:2013xfa}  that the 331 models with negative $\beta$ 
are most powerful in this case apply to $F_1$ representations. But as our 
analysis shows the model M2 with $\beta=-2/\sqrt{3}$ considered by us in 
\cite{Buras:2013dea} is disfavoured by the EWPO analysis.
\item
If the anomaly in question  remains but decreases with time also model M3 (left lower panel in Fig.~\ref{fig:ReC9vsBsmu}) and M14 (right lower panel in Fig.~\ref{fig:ReC9vsBsmuF2}) would be of interest.
\item
The remaining four models, in fact the four top models on our list of 
favourites in  (\ref{favoured}), do not provide any explanation of  
$B_d\to K^*\mu^+\mu^-$ anomaly but are interesting for $B_{s,d}\to\mu^+\mu^-$ 
decays. These are {\rm M6},  {\rm M8}, {\rm M9} and {\rm M11}, the first two 
with $F_1$ and the last two with $F_2$ fermion representation. The differences 
between these models as far as $B_s\to\mu^+\mu^-$ is concerned are most
transparently seen in Figs.~\ref{fig:BsmuRnu} and \ref{fig:BsmuRnuF2} from 
which we conclude that the strongest suppression of the rate for 
 $B_{s}\to\mu^+\mu^-$ can be achieved in M8 and M9.  M8 result is shown in the 
right upper panels in Figs.~\ref{fig:ReC9vsBsmu} and \ref{fig:BsmuRnu} and 
M9 result in the left upper panels of in Figs.~\ref{fig:ReC9vsBsmuF2} and \ref{fig:BsmuRnuF2}. In fact these two models are the two leaders on the list of 
favourites in  (\ref{favoured}).  The suppression of the $B_s\to\mu^+\mu^-$ rate is smaller in M6 and M11 
 as seen in lower right and lower left panels in Figs.~\ref{fig:ReC9vsBsmu} and \ref{fig:ReC9vsBsmuF2}, respectively.
\end{itemize}

\begin{figure}[!tb]
 \centering
\includegraphics[width = 0.44\textwidth]{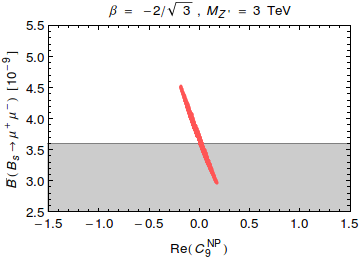}
\includegraphics[width = 0.44\textwidth]{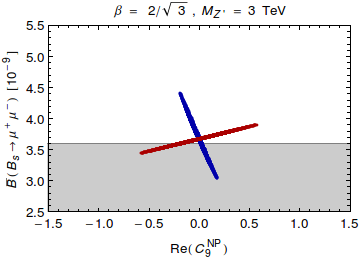}
\\
\includegraphics[width = 0.44\textwidth]{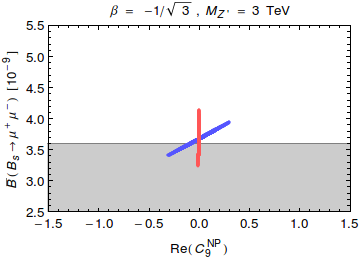}
\includegraphics[width = 0.44\textwidth]{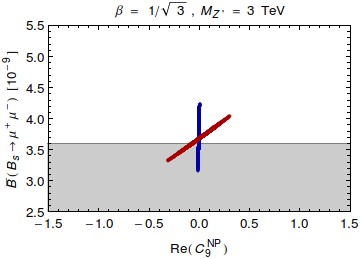}
\caption{\it $\overline{\mathcal{B}}(B_s\to\mu^+\mu^-)$ versus $\RE(C_9^\text{NP})$ for the favourite models from Tab.~\ref{tab:331pulls} for F1 (blue) and F2 (red) and
$\tan\bar\beta = 1(5)$ with lighter (darker) colours. 
}\label{fig:fav}~\\[-2mm]\hrule
\end{figure}

We observe that flavour physics can clearly distinguish between the 
favourite models selected by EWPO analysis. We summarize it in Fig.~\ref{fig:fav}, where only the results in the seven favourite models are shown. 
If the $B_d\to K^*\mu^+\mu^-$ anomaly will persist the winner will be M16 which  is represented in this figure by the dark red line in the upper right panel. If it disappears but suppression of 
the rate for  $B_{s}\to\mu^+\mu^-$ will be required the winners will be 
M8 and M9, the blue and red lines in the upper right and upper left panel 
in Fig.~\ref{fig:fav}, respectively. It is interesting that the combination of future flavour data 
and EWPO tests can rather clearly indentify one or two favourite 331 models 
among 24 cases considered by us.

An interesting situation would also arise if the LEP result for $\mathcal{A}_\ell$
would turn out to be the correct one. In this case M14 and M16 are no longer 
among favourites as seen in (\ref{favouredLEP}) and are replaced by 
M13 and M17. M13 is a good replacement for M14 as it also allows moderate 
softening of the $B_d\to K^*\mu^+\mu^-$ anomaly as seen in the 
right lower panel in Fig.~\ref{fig:ReC9vsBsmuF2}. But more interesting is 
M17. Indeed  as seen in the upper left panel in Fig.~\ref{fig:ReC9vsBsmu} 
for $\tan\bar\beta=0.2$ (gray line) the $B_d\to K^*\mu^+\mu^-$ anomaly can 
be softened as much as it was possible in the case of M16.  But as seen in 
the upper left panel in Fig.~\ref{fig:KCKMad} (purple line) in this model with 
$F_1$ representations also significant NP effects in $\klpn$ are possible. 
Smaller NP effects, as seen in  Fig.~\ref{fig:KCKMadF2}, are found  in the case of $F_2$ but $M18$ is not among the favourite models anyway. 
All this again shows an important interplay 
between flavour observables and electroweak precision tests.

\section{Summary and Conclusions}\label{sec:5}
In this paper we have addressed the question whether in 331 models the 
FCNCs due to $Z$ tree-level exchanges generated through $Z-Z^\prime$ mixing 
could play any significant role in flavour physics.
Actually it is known from the flavour analyses of 
Randall-Sundrum models with custodial protection (RSc) 
\cite{Blanke:2008yr,Bauer:2009cf}, that 
while $\Delta F=2$ processes are governed by heavy Kaluza-Klein gauge bosons with and without colour, NP contributions in  $\Delta F=1$ processes are governed 
by induced right-handed flavour-violating $Z$ couplings.

Here we analyzed  several 331 models which have a much smaller number of 
parameters than RSc and this allows to see more transparently various NP 
effects than in the latter scenario. As our basic formula for the 
$Z-Z^\prime$ in 331 model differs from the one found in the literature 
\cite{Diaz:2004fs,CarcamoHernandez:2005ka,Richard:2013xfa} we have reconsidered 
some aspects of constraints from EWPO. Moreover we have identified for the first time various correlations between flavour 
and electroweak physics that depend on the parameters of 331 model, in 
particular, $\beta$, $\tan\bar\beta$, $M_{Z^\prime}$ and the fermion representations. 

As far as flavour physics is concerned our main findings are as follows:
\begin{itemize}
\item
NP contributions to $\Delta F=2$ transitions and decays like $B\to K^*\ell^+\ell^-$ are governed by $Z^\prime$ tree-level exchanges. Therefore for these 
processes our analysis in \cite{Buras:2013dea} remains unchanged. But 
as we summarize below our analysis of EWPO casts some shadow on some 
of these models.
\item 
On the other hand for $B_{s,d}\to \mu^+\mu^-$ decays $Z$ contributions can 
be important. We find that for $\tan\bar\beta=5.0$ these contributions 
interfere constructively with $Z^\prime$ contributions enhancing NP 
 effects, while for low $\tan\bar\beta=0.2$ $Z$ contributions 
practically cancel the ones from $Z^\prime$. Similar dependence on 
$\tan\bar\beta$ is found for $K_L\to\mu^+\mu^-$.
\item
Similarly $Z$ boson tree-level contributions to $B_{s,d}$ and $K$ decays 
with neutrinos in the final state can be relevant but in this case 
the $\tan\bar\beta$ dependence is opposite to the one found for 
 $B_{s,d}\to \mu^+\mu^-$.  We find that for $\tan\bar\beta=5.0$ 
these contributions practically cancel the ones from $Z^\prime$ but
for low $\tan\bar\beta=0.2$ $Z$ contributions 
interfere constructively with $Z^\prime$ contributions enhancing NP 
 effects. In particular as seen in  Figs.~\ref{fig:KCKMad} and \ref{fig:KLCKMad}  in the case of $\klpn$ NP effects can amount to $30\%$ 
at the level of the branching ratio when the constraints from EWPO are not 
taken into account.
\item
As a result of this opposite dependence on $\tan \bar \beta$ the correlations between decays with muons and neutrinos in the final state exhibit significant 
dependence on $\tan\bar\beta$ and can serve to determine this parameter in 
the future. See in particular Figs.~\ref{fig:BsmuRnu} and \ref{fig:BsmuRnuF2}.
\item
Our analysis of $\epe$ is to our knowledge the first one in 331 models. 
Including both $Z^\prime$ and $Z$ contributions we find that the former 
dominate. But NP effects are not large and in order to fit the data 
 $\bsi\approx 1.0$ is favoured.
\item
We also find a strict correlation between   $\epe$ and $\mathcal{B}(\klpn)$. 
The interesting feature here, as seen in  Fig.~\ref{fig:Br0vsepsKCKMp}, 
is the decrease of $\epe$ with  increasing $\mathcal{B}(\klpn)$
 for {\it negative} $\beta$ and its increase with 
increasing $\mathcal{B}(\klpn)$  
 for {\it positive} $\beta$. 
\item
Performing the analysis for different fermion representations we find 
that for certain observables this dependence is significant. As an 
example the comparison of the plots in 
 Figs.~\ref{fig:ReC9vsBsmu} and \ref{fig:ReC9vsBsmuF2} 
demonstrates the breakdown of the invariance under (\ref{symmetry}) 
by $Z-Z^\prime$ mixing. 
\end{itemize}

As far as electroweak physics is concerned our main findings are as follows:
\begin{itemize}
\item
Seven among 24 combinations of $\beta$, $\tan\bar\beta$ and fermion representation $F_1$ or $F_2$ provide better or equally good description of the electroweak 
precision data compared with  the SM. However, none of these models allows for the explanation of the $2.5\sigma$ departures of $\mathcal{A}^b_{\rm FB}$ and
 $R_b$  from the SM  but several of them improve significantly the agreement of the theory with the average of SLD and LEP data for $A_l$. 
\item
Among these models none of them allows to simultaneously suppress the rate 
for $B_s\to\mu^+\mu^-$ and soften the $B_d\to K^*\mu^+\mu^-$ anomaly.
\item
On the other there are few models which either suppress the 
rate for $B_s\to\mu^+\mu^-$ or soften the $B_d\to K^*\mu^+\mu^-$ anomaly.
\item
None of these models allows significant NP effects in $B$ and $K$ decays 
with neutrinos in the final state although departures by $15\%$ relative 
to the SM prediction for the rate of $\klpn$ are still possible.
\item
Assuming that the LEP result for $\mathcal{A}_\ell$ is the correct one, we 
have found that in this case NP effects in $\klpn$ are larger than 
when both LEP and SLD results are taken into account.
\end{itemize}

Our analysis shows that the interplay of flavour physics and EWPO tests 
can significantly constrain NP models, in particular those with not too 
many free parameters. We are looking forward to coming years in order 
to see whether the 331 models will survive improved flavour data and in 
particular whether $Z^\prime$ will be discovered at the LHC. The 
correlations presented by us should allow to monitor transparently 
future developments in the data.

\section*{Acknowledgements}
We would like to thank Christoph Niehoff  and David Straub for asking us about 
the size of tree-level $Z$ contributions to FCNC processes in 331 models. 
We also thank Guido Altarelli, Jernej Kamenik, Ulrich Nierste, Francois Richard,  R. Martinez and F. Ochoa for interesting 
discussions.
This research was done and financed in the context of the ERC Advanced Grant project ``FLAVOUR''(267104) and was partially
supported by the DFG cluster
of excellence ``Origin and Structure of the Universe''.

\bibliographystyle{JHEP}
\bibliography{allrefs}
\end{document}